%% file: _PaperClimEconMeth_arxiv.tex
\documentclass[12pt]{article}
\usepackage{graphicx} 
\usepackage{xcolor}
\usepackage{amsmath}
\usepackage{amssymb}
\usepackage{hyperref} 
\usepackage{enumitem} 
\usepackage[round]{natbib}
\usepackage{doi}
\usepackage{cleveref}
\usepackage{booktabs, caption, colortbl, array, anyfontsize, multirow}
\usepackage{helvet}
\usepackage{datatool}
\usepackage[a4paper, margin=2.5cm]{geometry}
\usepackage{setspace}
\usepackage{authblk}

\newbox{\myorcidthanksbox}
\sbox{\myorcidthanksbox}{\large\includegraphics[height=1.8ex]{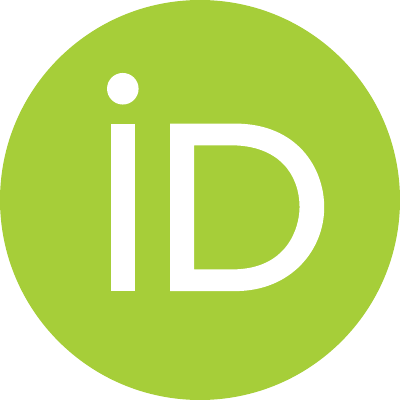}}
\newcommand{\orcidthanks}[1]{%
    \href{https://orcid.org/#1}{\raisebox{-0.5ex}{\usebox{\myorcidthanksbox}}\,#1}}

\author[1,2]{Christof Schötz\thanks{christof.schoetz@tum.de, \orcidthanks{0000-0003-3528-4544}, corresponding author}}
\author[2]{Jan Hassel\thanks{jan.hassel@pik-potsdam.de, \orcidthanks{0009-0004-9648-8326}}}
\author[2]{Christian Otto\thanks{christian.otto@pik-potsdam.de, \orcidthanks{0000-0001-5500-6774}}}
\affil[1]{Technical University of Munich, Germany; TUM School of Engineering and Design, Department of Aerospace \& Geodesy}
\affil[2]{Potsdam Institute for Climate Impact Research, Germany}

\title{Rethinking Climate Econometrics:\\Data Cleaning, Flexible Trend Controls, and Predictive Validation}
\date{}

\newcommand{\mo}[1]{\mathbf{#1}}
\newcommand{\mc}[1]{\mathcal{#1}}
\newcommand{\ms}[1]{\mathsf{#1}}
\newcommand{\mb}[1]{\mathbb{#1}}
\newcommand{\br}[1]{\left(#1\right)}
\newcommand{\brOf}[1]{\!\br{#1}}
\newcommand{\tr}{\!^\top\!}
\newcommand{\R}{\mathbb{R}}

\newcommand{\Reg}{\mathcal{S}}
\newcommand{\reg}{s}
\newcommand{\Tim}{\mathcal{T}}
\newcommand{\tim}{t}
\newcommand{\Ind}{\mathcal{I}}
\newcommand{\absof}[1]{|#1|}
\newcommand{\pr}{^\prime}
\DeclareMathOperator*{\argmin}{arg\,min}
\newcommand{\vn}[1]{\ensuremath{\mathsf{#1}}}

\newcommand{\loadmap}[2]{%
	\DTLloaddb{#1}{#2}%
	\DTLforeach{#1}{\Key=Key,\Value=Value}{%
		\expandafter\def\csname#1\Key\expandafter\endcsname\expandafter{\Value}%
	}%
}
\newcommand{\key}[2]{\csname #1#2\endcsname}
\loadmap{clean}{key/cleanCount.csv}
\loadmap{isimip}{key/isimipGrowthPrecentiles.csv}
\loadmap{tcReg}{key/tcReg.csv}
\loadmap{corr}{key/08_correlations_isimip_clean_log2growth_main.csv}
\loadmap{data}{key/04_dataInfo.csv}

\newcommand{\supplementref}[1]{Supplement #1}

\begin{document}
\maketitle
\vspace*{-0.5cm}
\noindent
\textbf{Funding:} J.H.\ and C.O.\ acknowledge funding from the European Union’s Horizon Europe research and innovation program under grant agreement no.~101081358 (ACCREU)

\noindent
\textbf{Acknowledgment:} The authors gratefully acknowledge Simon Philippssohn for his valuable contributions to the classification of economic outliers.

\noindent
\textbf{Data and Code:} All data and code necessary to reproduce the results are publicly available at \url{https://doi.org/10.5281/zenodo.15470558}.

\noindent
\textbf{Conflict of Interest Statement:} The authors declare no conflict of interest.

\begin{abstract}
    We assess empirical models in climate econometrics using modern statistical learning techniques. 
    Existing approaches are prone to outliers, ignore sample dependencies, and lack principled model selection. 
    To address these issues, we implement robust preprocessing, nonparametric time-trend controls, and out-of-sample validation across 700+ climate variables. 
    Our analysis reveals that widely used models and predictors—such as mean temperature—have little predictive power. 
    A previously overlooked humidity-related variable emerges as the most consistent predictor, though even its performance remains limited. 
    These findings challenge the empirical foundations of climate econometrics and point toward a more robust, data-driven path forward.
\end{abstract}
\input{sec_intro.tex}

\input{sec_setup.tex}
\input{sec_clean.tex}
\input{sec_fe.tex}
\input{sec_selection.tex}
\input{sec_results.tex}
\input{sec_discussion.tex}
\bibliography{ref.bib}
\renewcommand{\thefigure}{S\arabic{figure}}
\renewcommand{\thetable}{S\arabic{table}}
\renewcommand{\theequation}{S\arabic{equation}}
\renewcommand{\thesection}{S\arabic{section}}
\setcounter{section}{0}
\setcounter{table}{0}
\setcounter{equation}{0}
\setcounter{figure}{0}
\clearpage
\part*{Supplement for:\\Rethinking Climate Econometrics:\\Data Cleaning, Flexible Trend Controls, and Predictive Validation}
\clearpage
\input{app_details.tex}

\clearpage
\bibliography{ref.bib}
\end{document}

%% file: sec_intro.tex
\section{Introduction}

Climate econometrics is the study of how weather, climate, and climate change affect economic outcomes using statistical methods. Understanding these relationships is essential for informing policy, managing climate-related risks, and guiding adaptation strategies. As interest in the economic implications of climate variability and change has grown, the field has developed at the intersection of economics, climate science, and statistics \citep{dell_what_2014,hsiang_climate_2016}.

A widely used empirical approach in climate econometrics is fixed effects panel data regression, where macroeconomic indicators—such as a country’s GDP—are regressed on climate variables like average annual temperature. This method aims to quantify the extent to which economic variation can be statistically attributed to fluctuations in climatic conditions. Fixed effects account for time-invariant country characteristics and global shocks shared across countries, while additional controls, such as country-specific time trends, are often included to capture heterogeneous growth trajectories.

Numerous influential studies have adopted this methodology using global country-year panel datasets. For example, \citet{dell_temperature_2012} estimate linear effects of temperature and precipitation on economic output, finding that higher temperatures negatively affect economic growth in poorer countries. \citet{burke_global_2015} extend this framework by including quadratic terms for temperature and precipitation as well as quadratic time trends, uncovering a significant nonlinear (inverted U-shaped) relationship between temperature and GDP growth. \citet{pretis_uncertain_2018} incorporate measures of monthly variance and extreme temperatures, along with dummy variables for outlier events, to assess differences in projected economic damages under 1.5\textdegree C and 2\textdegree C warming scenarios.
Other contributions have expanded the analysis by increasing the spatial resolution of the panel. \citet{kalkuhl_impact_2020} use economic data of subnational regions to explore temperature effects. Continuing the analysis of subnational data, \citet{kotz_day--day_2021} investigate the role of intra-annual temperature variability, while \citet{kotz_effect_2022} examine the effects of diverse precipitation-related indicators. \citet{newell_gdp-temperature_2021} assess model performance through out-of-sample testing, finding no significant predictive power of temperature on GDP growth, though more robust effects are observed for GDP levels. \citet{kahn_long-term_2021} study deviations from long-term temperature norms, and \citet{krichene_social_2023} incorporate extreme weather events to examine longer-term macroeconomic consequences.

Several meta-analyses and review articles provide a broader synthesis of this literature and the methodologies it employs \citep{dell_what_2014, hsiang_climate_2016, howard_few_2017, nordhaus_survey_2017, kolstad_estimating_2020, chang_temperature_2023, intergovernmental_panel_on_climate_change_ipcc_climate_2023, tol_meta-analysis_2024}.

While fixed effects panel regression is a standard tool in applied econometrics \citep{arellano_panel_2010,wooldridge_econometric_2010}, its limitations have received increasing attention. \citet{hill_limitations_2020} and \citet{imai_use_2021} emphasize that the validity of such models hinges on strong assumptions about functional form, treatment homogeneity, and independence, which are often untested in practice. Violations of these assumptions can create unreliable and misleading findings.
Although techniques such as out-of-sample testing and formal variable selection are standard in other empirical disciplines, they are rarely applied in the current climate econometrics literature. This limits the robustness and generalizability of existing findings. In this article, we address several of these methodological gaps and present results based on more rigorous, data-driven approaches.

We highlight the problematic influence of outliers in economic data and provide a cleaned dataset that enhances the robustness of our results. To address temporal confounding, we introduce nonparametric time trend controls that flexibly account for slowly evolving, country-specific characteristics. These controls avoid the arbitrary specification of polynomial time trends and their induced long-range serial correlation biases. In addition to examining temporal dependencies, we assess spatial correlation and its implications for standard error estimation.

In our main analysis, we apply various statistical learning methods for variable selection to identify the most predictive climate indicators of GDP growth and levels from a pool of over 700 variables (roughly $10^{213}$ possibilities of predictor combinations) and test model performance in an out-of-sample framework. Among the methods, penalized regression techniques—LASSO, Ridge regression, and Elastic Net \citep{Hastie2009}—consistently yield the best predictive accuracy. The most robust predictor we identify is the squared maximum specific humidity of the previous year, which may influence economic activity through channels such as wet-bulb temperatures or extreme precipitation events. In contrast, commonly used variables such as mean annual temperature and its square show no predictive power, calling into question some earlier findings in the literature. Overall, even the best-performing models exhibit limited predictive power. We identify a lack of temporal stationarity as a possible reason for failings to be predictive out-of-sample.

%% file: sec_setup.tex
\section{Setup}\label{sec:general}
We first provide a description of the basic statistical framework, the panel data regression, and the data and variables used in this framework.

\subsection{Panel Data Regression with Fixed Effects}\label{sec:setup:general}

Our goal is obtain empirical evidence of the influence of climate variables $\mo X_1, \dots, \mo X_m$ on a target economic variable $\mo Y$.
All variables are indexed by two dimension: One is spatial (here: countries), denoted by $\reg\in \Reg$. The other one is temporal (here: years), denoted by $\tim\in\Tim$, where $\Tim \subseteq \mb Z$ is a set of consecutive time points. Thus, the data forms a panel and panel data analysis methods can be applied. We do not assume that the panel is full, i.e., we may only observe certain region--time combinations $(\reg, \tim) \in \Ind$ for some index set $\Ind \subseteq \Reg\times \Tim$. Denote the total number of observations as $n = \# \Ind$.

To achieve our goal, we apply a linear regression model of the form
\begin{equation}\label{eq:base:expanded}
	Y_{\reg,\tim} = Z_{1,\reg,\tim} \beta_1 + \dots + Z_{p,\reg,\tim} \beta_p + w_{1,\reg,\tim} \alpha_1 + \dots +  w_{q,\reg,\tim} \alpha_q + \varepsilon_{\reg,\tim}, \quad (\reg,t)\in \Ind.
\end{equation}
Here $Z_{k,\reg,\tim}$ are features derived from the climate variables. These can, e.g., be transformations like $\log(X_{j,\reg,\tim})$ or $X_{j,\reg,\tim}^2$; or time lags like $X_{j,\reg,\tim-1}$, $X_{j,\reg,\tim-2}$, ...; or interaction terms like $X_{j,\reg,\tim}X_{j\pr,\reg,\tim}$.
The terms $w_{k,\reg,\tim}$ are fixed effect variables, e.g., dummy variables $w_{k,\reg,\tim} = 1_{\tim = \tim_0}$ for a given time point $\tim_0\in\Tim$ or time trends $w_{k,\reg,\tim} = t 1_{\reg = \reg_0}$ for a given region $\reg_0\in\Reg$.
The parameters $\alpha_j$, $\beta_j$ are unknown. They differ in that we are interested in statistical inference on $\beta_j$, whereas $\alpha_j$ are nuisance parameters that model unobserved effects that we want to control for but have no further interest in.
The errors $\varepsilon_{\reg,\tim}$ are modeled as centered random variables, which are normally distributed or fulfill moment conditions that allow to use asymptotic normality in later analysis.
In matrix notation, we have
\begin{equation}\label{eq:base:matrix}
	\mo Y = \mo Z \beta + \mo w \alpha + \varepsilon
\end{equation}
where
\begin{align}
	\mo Y &= ( Y_{\reg,\tim})_{(\reg,\tim)\in\Ind}\in\R^n, &
	\beta &= (\beta_1, \dots, \beta_p)\in\R^p,\\
	\mo Z &=  (Z_{j,\reg,\tim})_{(\reg,\tim)\in\Ind, j=1,\dots,p}\in\R^{n\times p},&
	\alpha &= (\alpha_1, \dots, \alpha_q)\in\R^q,\\
	\mo w &=  (w_{k,\reg,\tim})_{(\reg,\tim)\in\Ind, k=1,\dots,q}\in\R^{n\times q},&
	\varepsilon &= (\varepsilon_{\reg,\tim})_{(\reg,\tim)\in\Ind}\in\R^n.
\end{align}
Aside from statistical inference on $\beta$, we are also interested in model selection or feature selection, i.e., finding a set of features $Z_{k,\bullet,\bullet}$ that yields a model with predictive power for unseen observations, i.e., a model that generalizes well. Higher predictive power can be evidence for causal relationships, which supports the use of a model in future projections. In order to find such a model, we need the set of selected features to be as small as possible while explaining as much as possible of the variation of the target $Y_{\bullet, \bullet}$. Model selection is explored in \cref{sec:modelselection}.
\subsection{Data}
Our primary data source is the Inter-Sectoral Impact Model Intercomparison Project (ISIMIP). Specifically, we utilize climate variables from the GSWP3-W5E5 dataset \citep{isimip_clim}, the tropical cyclone data \citep{isimip_tc}, as well as GDP (PPP) \citep{isimip_gdp} and country-level population data \citep{isimip_pop}. In addition, we incorporate gridded population data from the History Database of the Global Environment (HYDE), version 3.3 \citep{hyde3p3}. To assess the robustness of our findings with respect to economic data sources, we also employ World Development Indicator GDP data \citep{worldbank2025gdp}. All datasets are restricted to the period \key{data}{yearMin}--\key{data}{yearMax}, and our analysis covers a total of \key{data}{nCountry} countries.

\begin{table}[h!]
	\begin{center}
		\fontsize{8.0pt}{10pt}\selectfont
		\fontfamily{phv}\selectfont
		\renewcommand{\arraystretch}{1.05}
		\setlength{\tabcolsep}{0.3em}
		\rowcolors{2}{gray!20}{white}
		\begin{tabular}{ll}
			\toprule
			\textbf{Variable} & \textbf{Description} \\
			\midrule\addlinespace[2.5pt]
			hurs & Near-surface relative humidity \\
			huss & Near-surface specific humidity \\
			pr & Total precipitation \\
			prsn & Snowfall \\
			ps & Surface air pressure \\
			rlds & Long wave downwelling radiation \\
			rsds & Short wave downwelling radiation \\
			sfcwind & Near-surface wind speed \\
			tas & Daily mean temperature \\
			tasmax & Daily maximum temperature \\
			tasmin & Daily minimum temperature \\
			\bottomrule
		\end{tabular}
	\end{center}
	\caption{Climate-related base variables from GSWP3-W5E5 and their descriptions.}
	\label{tbl:climvars}
\end{table}
\subsection{Climate Econometric Model}
Our target variable in the regression is the economic growth rate $Y_{\reg,\tim} = \log(\vn{GDPpc}_{\reg,\tim}/\vn{GDPpc}_{\reg,\tim-1})$, where $\ms{GDPpc}_{\reg,\tim}$ is the gross domestic product per capita in country $\reg$ and year $\tim$ either using the ISIMIP or the World Bank data.

Most of the predictor variables we use, are based on the 11 climate variables given in \cref{tbl:climvars}. All variables are first given as daily gridded ($0.5^{\circ} \times 0.5^{\circ}$) values. For each grid cell, they are aggregated to yearly values using the mean, standard deviation, minimum, and maximum. This results in 44 variables indexed by grid cell and year. We have additional 12 variables derived from \vn{tas} and \vn{pr} (\cref{tbl:climvars}) by calculating the deviation from a rolling window of 11, 31, and 91 days (\vn{prDevi11}, \vn{prDevi31}, \vn{prDevi91}, \vn{tasDevi11}, \vn{tasDevi31}, \vn{tasDevi91}) and the rate of extreme events larger than $1-2^{-8} \approx 99.6\%$, $1-2^{-10} \approx 99.9\%$, $1-2^{-12} \approx 99.98\%$ of the whole observation time distribution (\vn{prExtr08}, \vn{prExtr10}, \vn{prExtr12}, \vn{tasExtr08}, \vn{tasExtr10}, \vn{tasExtr12}).
All yearly gridded values are aggregated to country means using population as grid weights.
Furthermore, we add the number people affected by tropical cyclones of at least 34 knots, 48 knots and 64 knots (\vn{tc34kt}, \vn{tc48kt}, \vn{tc64kt}) to our mix of variables. We arrive at $44 + 12 + 3 = 59$ variables.

For each of the 59 base variables, we create linear and squared terms and up to 5 lags. This leaves us at $59 \cdot 2 \cdot 6 = 708$ predictors to choose from. Furthermore, we have all variables in a base form ($Z_{k,\reg,\tim}$) and a difference form ($\Delta Z_{k,\reg,\tim} = Z_{k,\reg,\tim} - Z_{k,\reg,\tim-1}$), leading to two different kinds of regression models. Using differenced variables is sometimes called level regression as opposed to growth regression which uses the original variables. In both cases, we use the same target $Y_{\reg,\tim}$ as described before, which can be written as $Y_{\reg,\tim} = \Delta\log(\ms{GDPpc}_{\reg,\tim})$. 
We do not mix original and differenced predictors within the same model and do not delve into the interpretation of the two modeling strategies here. Instead, we refer readers to \citet{newell_gdp-temperature_2021} for a detailed discussion and present results for both approaches.

%% file: sec_clean.tex
\section{Step 1: Cleaning the Data}\label{sec:clean}
Ensuring the quality of the data set is a critical first step before performing regression analysis. This involves identifying and addressing outliers and out-of-distribution data that could bias results or compromise the validity of statistical inference. By systematically removing problematic data points, we aim to create a robust foundation for accurate modeling and reliable inference.
\subsection{Outliers}\label{sec:clean:out}
Outliers in the data can significantly distort results, potentially overshadowing true effects or introducing spurious findings that lack causal basis. Broadly speaking, outliers are data points that appear highly atypical or unexpected relative to the statistical model. We categorize outliers into two primary types here: mistakes and nuisance events.

Mistakes refer to incorrect data points, often due to issues such as instrument failure or transcription errors. As an example, version 2 of the DOSE data set \citep{dose} showed a decline in economic production per capita in Madrid, Spain, of almost 90\% from 1994 to 1995. In version 4, there is instead an increase by 5\%. The steep decline in the old version might be due to a wrongly placed decimal point. See \supplementref{Table S6} for more details on this example. Another mistake is the \vn{GDPpc} in Oman in 1965 in the World Bank data, where it is reported to grow by a factor of $10^6$ (by many orders of magnitude the largest value in the dataset and clearly wrong).\footnote{We reported this issue on February 14, 2025, which lead to a correction of this mistake in a subsequent version of the data.}

In contrast, nuisance events are data points that accurately reflect reality but arise from rare, extreme non-climate events, which are not captured by the model and yet can be understood through external knowledge sources. For example, the ISIMIP economic data shows a decline of economic production per capita in Iraq in the year 1991 by 65\%. This is an extreme event in the data compared to 90\% of the data having growth values between \key{isimip}{P05} and \key{isimip}{P95}. Furthermore, it can be attributed to a non-climate event: the Gulf War. See \supplementref{Table S8} for more details on this example.
\subsection{Off-Distribution Data}\label{sec:clean:off}
Another type of data that can disrupt a statistical analysis is \emph{Off-Distribution Data}, which refers to data that does not originate from the distribution under study. This type of data often enters a dataset through processes like imputation, where missing values are filled with proxy estimates derived from other available information instead of direct measurements. While such data can be valuable for maintaining dataset completeness, it may also significantly skew the results of a statistical analysis. Unlike outliers, Off-Distribution data points typically have values that align with the general range of the target distribution, making them less conspicuous. However, they often exhibit distinct statistical properties, such as reduced variance or artificially inflated correlations with other variables. These discrepancies can introduce bias or lead to invalid conclusions in the analysis.

For instance, consider the ISIMIP GDP data for post-Soviet states, which spans from 1960 to 2021, including periods before these states existed independently of the Soviet Union. In the dataset, economic growth data for the years 1960 to 1985 is imputed with a linear method, see \supplementref{Figure S1}. This imputation introduces an artifact where the correlation coefficient between any two post-Soviet states for the economic growth rate equals $1$. Consequently, there is no independent statistical information about economic changes of individual states during this period beyond what is reflected in the aggregated output of the Soviet Union. See \cref{sec:clean:offfind} for a more comprehensive correlation analysis.
\subsection{Example of the Impact of Outliers}\label{sec:clean:outexa}
As an example, we use the following model to find the economic impact of tropical cyclone and changes in mean annual surface temperature:
\begin{equation}\label{eq:model:tc}
		Y_{\reg,t}
		=
		\beta_{\ms{tas}}X_{\ms{tas},\reg,t}
		+ \beta_{\ms{tas^2}}X_{\ms{tas},\reg,t}^2
		+ \sum_{\ell=0}^{5} \beta_{\ms{tc},\ell}X_{\ms{tc},\reg,t-\ell}
		+
		\alpha_t
		+ \alpha_{\reg, 0} + \alpha_{\reg, 1} t + \alpha_{\reg, 2} t^2
		+ \varepsilon_{\reg,t}
		\,,
\end{equation}
where, for region $\reg$ and year $t$, $Y_{\reg,t} = \log(\vn{GDPpc}_{\reg,t}/\vn{GDPpc}_{\reg,t-1})$, $X_{\ms{tas},\reg,t}$ is the mean (population weighted) annual near-surface air temperature (tas), and $X_{\ms{tc},\reg,t-\ell}$ is the mean ratio of people affected by a tropical cyclone (TC) with $\ell$ lag years.

Our initial results indicate that lags $\ell=1,3,5$ of tropical cyclone events are significant, while temperature predictors are not. Through influence analysis, we identify the 2002 observation from the U.S.\ Virgin Islands (VIR) as exerting the greatest influence on the lag terms with a growth value of $+48.7\%$, which ranks as the $11$th highest out of \key{tcReg}{nRaw} observations. This value seems to be an artifact of the dataset (likely a change in source data between 2001 and 2002\footnote{The World Bank GDP time series for U.S.\ Virgin Islands starts at 2002} rather than a real event\footnote{According to the US Virgin Islands Bureau of Economic Research, \emph{there was some weakening in the performance of the US Virgin Islands economy during 2002}, see \url{https://usviber.org/wp-content/uploads/2022/07/EconReviewJan2003.pdf}}. Removing this single influential observation changes the statistical significance of the lagged effects strongly, and with all outliers and off-distribution data excluded, surface air temperature and its quadratic term become significant. See \cref{tbl:tcRegressionExapleOutliers} for details.

\begin{table}
	\input{tbl/05_tableTcOutliers.tex}
	\caption{Coefficients and $p$-values for applying model \eqref{eq:model:tc} to different subsets of the ISIMIP dataset. \emph{Raw} is the whole dataset; for \emph{Clean 1}, VIR 2002 is removed; for \emph{Clean All}, the cleaned economic data as described in \cref{sec:clean:cleaned} is used.}\label{tbl:tcRegressionExapleOutliers}
\end{table}
\subsection{Treating Outliers}\label{sec:clean:outtreat}
A straightforward approach to detecting outliers involves inspecting the largest and smallest values of each variable; unusually high or low values may indicate potential outliers. In connection with further knowledge about the data, we can identify outliers. For example \cref{tbl:extremeGdppcChange} shows the largest and smallest GDP growth values in the ISIMIP dataset. The most decline in GDP is observed in Iraq year 1991. This extreme value can be attributed to the Gulf War and thus be counted as an outlier (it is extreme and not climate related).

\begin{table}
	\input{tbl/01_isimip_gdppc_extreme10.tex}
	\caption{For the ISIMIP economic data, the most extreme changes in $\vn{GDPpc}$ as identified by $\Delta\vn{GDPpc}_{\reg, t} = (\vn{GDPpc}_{\reg,t}-\vn{GDPpc}_{\reg,t-1}) / \vn{GDPpc}_{\reg,t-1}$. For a translation of the ISO country code to a country name, see \supplementref{Table S9 to S11}. Less than 1\% of all 15,189 growth values in the dataset are below $-12\%$ and less than 1\% are above $14\%$.}\label{tbl:extremeGdppcChange}
\end{table}

In this study, we manually review potential outliers characterized by extreme growth events and investigate their causes through online searches. If a non-climate-related explanation for the extreme growth is identified, we remove the corresponding data point from the dataset. The resulting dataset is described in \cref{sec:clean:cleaned}.

It is crucial to emphasize that we do not begin by performing a regression analysis to identify points with high residuals or those that might contradict our hypothesis. Such an approach could risk (either intentionally or unintentionally) introducing $p$-hacking. Instead, we base the exclusion of data points on external, independent information, ensuring that the removal decision is not influenced by its effect on the regression results.
\subsection{Alternative Treatments of Outliers}\label{sec:clean:alt}
Additionally, external datasets can be utilized to supplement information when a given data point is deemed unreliable. For instance, one might use a conflict dataset, such as the \emph{Uppsala Conflict Data Program} \citep{ucdp}, to exclude data points associated with armed conflicts of a certain scale. While this approach has the merits of a more formal and less laborious procedure than checking extreme values by hand, it is not without limitations. In particular, the time frame or the magnitude of the conflict and the economic signal too often do not align. Overall, although such datasets can provide valuable insights, the results lack sufficient precision, leading us to forgo this approach.

Another alternative approach is robust regression, which automatically down-weights data points that deviate significantly from the model. However, this method has a key drawback: extreme weather events, which often produce such outlier data points, are precisely the phenomena we aim to analyze. Excluding or down-weighting these points would mean losing valuable information, even if retaining them increases variance and uncertainty in the results. Additionally, applying robust regression to a large dataset with many predictors can be computationally expensive and numerically unstable. See \supplementref{section S3} for an example.

Instead of removing outliers, one could introduce a dummy variable, such as one indicating the presence of armed conflict. However, because the economic consequences of armed conflict vary significantly across cases, this approach would not fully address the outlier characteristics of the affected data points.

Alternatively, one might choose to ignore outliers altogether, assuming they are neither numerous nor severe enough to substantially influence the results. However, this approach is only valid if an analysis demonstrates that the impact of outliers is indeed negligible. As shown above, outliers can have a significant effect on the results, making this a risky assumption.
\subsection{Finding and Treating Off-Distribution Data}\label{sec:clean:offfind}
We focus on identifying off-distribution data in the economic target variable of the ISIMIP dataset. To achieve this, we perform two key analyses: a correlation-based clustering of time series to detect unusual patterns and a search for constant segments within the data. Based on these analyses, we identify and remove off-distribution observations.

Specifically, we compute the correlation matrix between each pair of economic time series of the form
$((Y_{\reg,\tim})_{\tim=\tim_{\ms{from}}, \dots, \tim_{\ms{to}}}, (Y_{\reg',\tim})_{\tim=\tim_{\ms{from}}, \dots, \tim_{\ms{to}}})$, $\reg\neq \reg'$, where $Y_{\reg,\tim}$ is the economic growth rate for country $\reg$ in year $\tim$. Using these correlations, we perform hierarchical clustering\footnote{We employ the unweighted pair group method with arithmetic mean (UPGMA).} to identify clusters of highly correlated countries.

Since imputation is often applied to segments of time series data, we compute correlations for each time interval $[t_{\ms{from}}, t_{\ms{to}}]$ within the dataset's full time range, ensuring a minimum interval length of $7$ years ($t_{\ms{to}} - t_{\ms{from}} \geq 7$) to obtain reliable results. Additionally, to minimize false positives due to spurious correlations, we only consider clusters with extremely high mean correlations. The resulting clusters, with mean correlations close to $1$, are shown in \supplementref{Table S12}.

The ISIMIP dataset also includes time series with segments of exactly constant economic output, which are likely the result of imputation. These segments, having zero variance, produce undefined correlations with other variables, making correlation analysis inapplicable. To address this, we identify all segments where the time series remains exactly constant for at least $3$ years (zero growth over at least $2$ years) and mark them as off-distribution. These segments are detailed in \supplementref{Table S13}.

Using these findings, along with additional historical context, we identify and remove observations classified as off-distribution, as described in \cref{sec:clean:cleaned}.
\subsection{A Cleaned Dataset}\label{sec:clean:cleaned}
To obtain a cleaned economic dataset from the original ISIMIP data, we have marked: \key{clean}{outlier} observations as (non-climate related) outliers by checking the most extreme growth events individually; \key{clean}{offCorrelated} observations as imputed because of highly correlated time series segments; \key{clean}{offZeroGrowth} observations as imputed because of constant time series segments. Additionally, we remove post-Soviet and post-Yugoslavia observations up to 1996 (\key{clean}{offHistorical} observations), as they are partially imputed and partially reflect political changes so that we deem them uninformative on climate events. From the original \key{clean}{raw} values, we remove \key{clean}{total} observations in total, which is less than the sum of the number of elements in the previously mentioned categories as those intersect. Thus, the final dataset has \key{clean}{clean} values.

%% file: sec_fe.tex
\section{Step 2: Controlling for Global Shocks and Regional Time Trends}\label{sec:fe}
To find evidence for a causal connection between economic and climate variables, we have to control for time-point specific and region specific effects. To illustrate, let us consider the following version of \eqref{eq:base:expanded},
\begin{equation}\label{eq:femodel}
	Y_{\reg,\tim} = f_\beta(Z_{\bullet,\reg,\tim}) + \alpha_{\tim} + g_{\reg}(\tim) + \varepsilon_{\reg,\tim}\,.
\end{equation}
Here $f_\beta$ is a function that models the influence of the predictors on the response; $\alpha_{\tim}$ is the global time-specific fixed effect (growth of the global economy in year $\tim$); $g_{\reg}(\tim)$ is the region specific time trend control (the smooth time trend of economic growth of a given region), e.g., $g_s(\tim) = \alpha_{\reg,0} + \alpha_{\reg,1} \tim$ for a linear time trend; and $\varepsilon_{\reg,\tim}$ models the remaining error of the model.

The fixed effect $\alpha_{\tim}$ accounts for time-specific shocks affecting all regions, such as global economic crises, while $g_{\reg}(\tim)$ captures region-specific long-term trends in economic growth, such as persistent differences between developing and developed countries.
\subsection{The Power of Frisch--Waugh--Lovell}
The number of control parameters $\alpha_{\tim}, \alpha_{\reg,0}, \alpha_{\reg,1}$ for $\reg\in\Reg,\tim\in\Tim$ may be large while the total number of observations may be to small to accurately estimate their values. Fortunately, the Frisch--Waugh--Lovell (FWL) theorem demonstrates that this limitation has minimal impact on the accuracy of the parameters of interest, $\beta$.

Recall the general regression setup introduced in \cref{sec:setup:general}. By the FWL theorem, the least squares estimate of the parameters of interest $\beta$ in the model
$\mo Y = \mo Z \beta + \mo w \alpha + \varepsilon$ is equivalent to the least squares estimate of $\beta$ in the transformed model
\begin{equation}
    \tilde{\mo Y} = \tilde{\mo Z} \beta + \tilde\varepsilon,
\end{equation}
where the transformed variables $\tilde{\mo Y}$ and $\tilde{\mo Z}$ are obtained by removing the components explainable by $\mo w \alpha$. Specifically, $\tilde{\mo Y} = M \mo Y$ and $\tilde{\mo Z} = M \mo Z$, with $M$ being the linear projection onto the complement of the column space of $\mo w$:
\begin{equation}
    M = I - \mo w (\mo w\tr \mo w)^{-1} \mo w\tr
\end{equation}
where $I \in \mathbb{R}^{n \times n}$ is the identity matrix.

This result not only ensures that $\beta$ can be accurately estimated without precise estimation of $\alpha$, but also offers significant computational advantages when analyzing panel data with many control variables.
Given the raw data $(X_{\bullet,\reg,\tim}, Y_{\reg,\tim})_{(\reg,\tim) \in \Ind}$, we can streamline the computation by first creating the features $Z_{\bullet,\reg,t}$ to analyze. Next, we apply a control variable design and transform the target and features to obtain the controlled dataset $(\tilde Z_{\bullet,\reg,\tim}, \tilde Y_{\reg,\tim})_{(\reg,\tim) \in \Ind}$. This preprocessing step decouples the estimation of \(\beta\) from the high-dimensional controls, significantly reducing computational complexity.

This approach is especially advantageous when the dimension $q$ of $\alpha$ is large relative to the dimension $p$ of $\beta$. For instance, in the example described in \eqref{eq:model:tc}, we have \key{clean}{nCountry} countries and \key{clean}{nYear} years, requiring $q = \key{clean}{nFe2}$ control parameters if quadratic time trends are included. In contrast, the parameter of interest, $\beta$, has dimension $p = 8$. While $q$ may still be manageable for a single regression, scenarios involving bootstrap or cross -alidation (as discussed in \cref{sec:modelselection}) often require thousands of regressions with the same controls. In such cases, the two-step procedure enabled by the FWL theorem provides immense computational benefits. Furthermore, covariance matrices needed for confidence intervals and hypothesis tests can be computed in the $p\times p$ dimensions of the transformed model instead of $(q+p) \times (q+p)$ dimensions of the original model equation (if small sample corrections are adjusted correctly).
\subsection{Local Time Trends}\label{ssec:localtimetrends}

To account for the long-term time trend of each country, the function $g_{\reg}(\tim)$ in \eqref{eq:femodel} must be specified. A common approach is to model $g_{\reg}(\tim)$ as a polynomial in $\tim$, with the degree typically ranging from 0 (constant, capturing only the country-specific mean growth) to 3 (cubic, allowing for up to two changes in direction). While low-degree polynomials capture only simple trends, higher degrees may inadvertently remove meaningful information that should be attributed to the parameters of interest.

\begin{figure}
	\vspace*{-1cm}
	\includegraphics[width=0.95\textwidth]{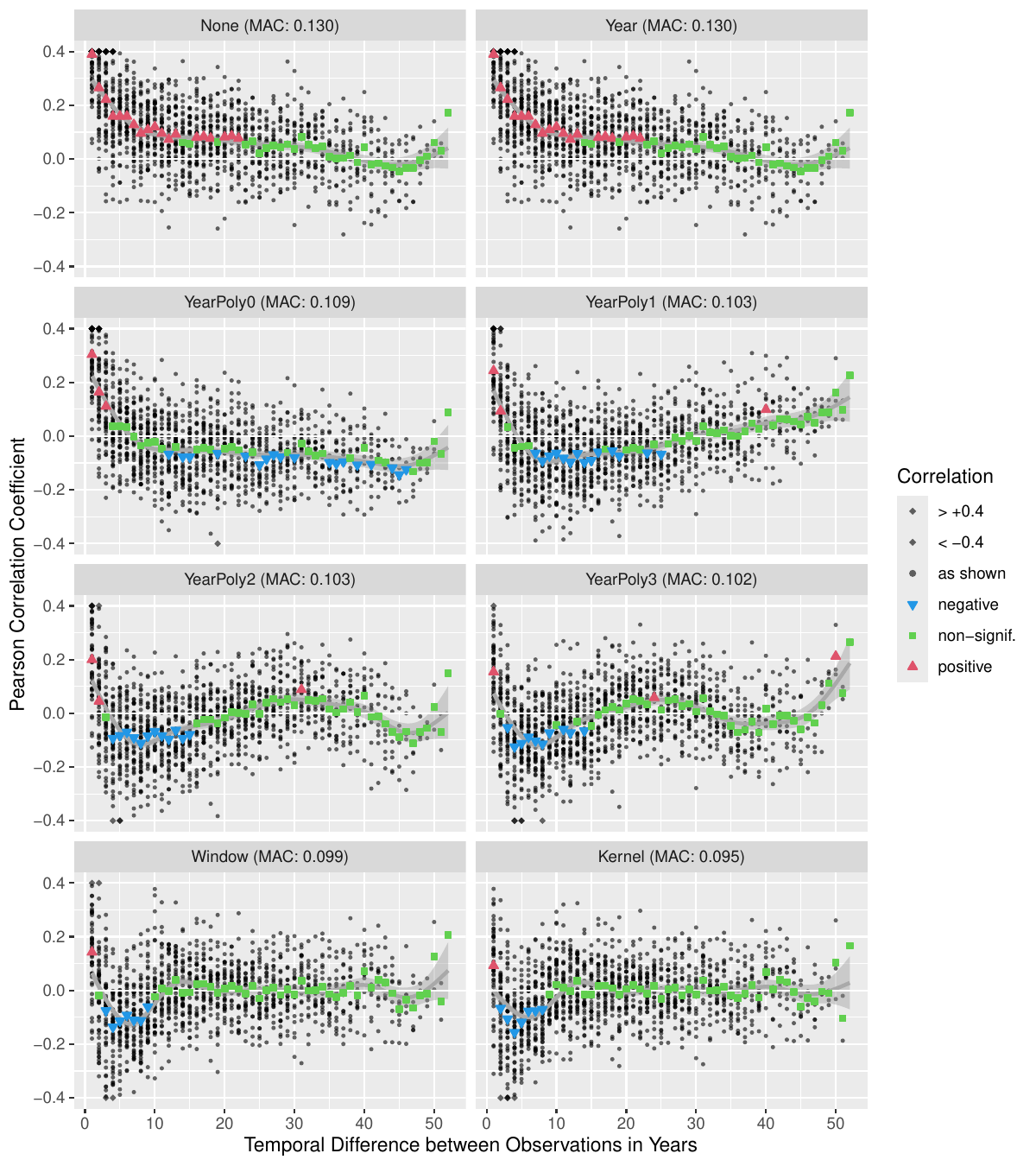}
	\caption{Effect of different control variable designs on temporal correlation of residual economic growth after controlling. Each black dot is an empirical correlation value between two different years calculated from the ISIMIP economic data. The gray curve shows a Loess smoother of these correlations. For each fixed temporal difference, we test whether the correlations values for pairs of years with this given difference are significantly different from zero via a conservative permutation test, see \supplementref{section S4}. Furthermore, we apply the Benjamini-Hochberg correction (with false discovery rate level of $0.05$) to account for the multiple testing issue. Blue and red triangles show correlations significantly different from zero, whereas green squares indicate that non-zero correlation cannot be detected. MAC is the mean absolute correlation. Without controls (\textit{None}), we observe a high positive correlation for more than 20 years. If only year fixed effects are applied, the result is exactly the same as this only changes the means of the correlated variables. All polynomial time trends are able to reduce the positive correlation of small temporal difference, but introduce negative correlations of temporal differences between 3 and 14 (and more) years. Both local control designs remove correlations in later years. Only (mostly negative) correlations remain for temporal differences of six to seven years. Overall the Kernel-based method seems to show the best results.}
	\label{fig:fe:corrtemp}
\end{figure}

However, polynomial time trends have notable drawbacks. First, practitioners must choose the polynomial degree, and there is no universally accepted rule for making this decision. This may (involuntarily) lead to $p$-hacking when the choice of polynomial degree is influenced by the desired claim of the research article. Second, even the best-fitting polynomial may fail to adequately approximate the time trend for all regions. Third, polynomials can also introduce artificial long-distance correlations, which may distort the results (see \cref{fig:fe:corrtemp}).

Rather than relying on polynomials, we propose a more flexible nonparametric approach that directly embodies the idea of a \textit{typical value} for a given time period—longer than a single time point but shorter than the entire time interval. Specifically, we use a local linear estimator for the time trend. We refer to this method as \textit{Window} and \textit{Kernel}, respectively, depending on whether all observations are weighted equally or according to a quadratic kernel. These local linear methods resemble a rolling window approach (local constant) but offer statistical advantages, particularly near the edges of the time interval \citep{fan1996local}.

As with the rolling window method, the local linear estimator requires a parameter to define how much of the surrounding data influences the estimate at each time point. This parameter, known as the bandwidth, determines the threshold for considering neighboring time points when calculating the trend. To make the method fully automatic, we propose using leave-one-out cross-validation with the economic response variable as target to select the optimal bandwidth. The automatically selected values are such that 10 to 12 years before and after a given year are used to calculate the local time trend. 

\begin{figure}
	\vspace*{-1cm}
	\includegraphics[width=0.95\textwidth]{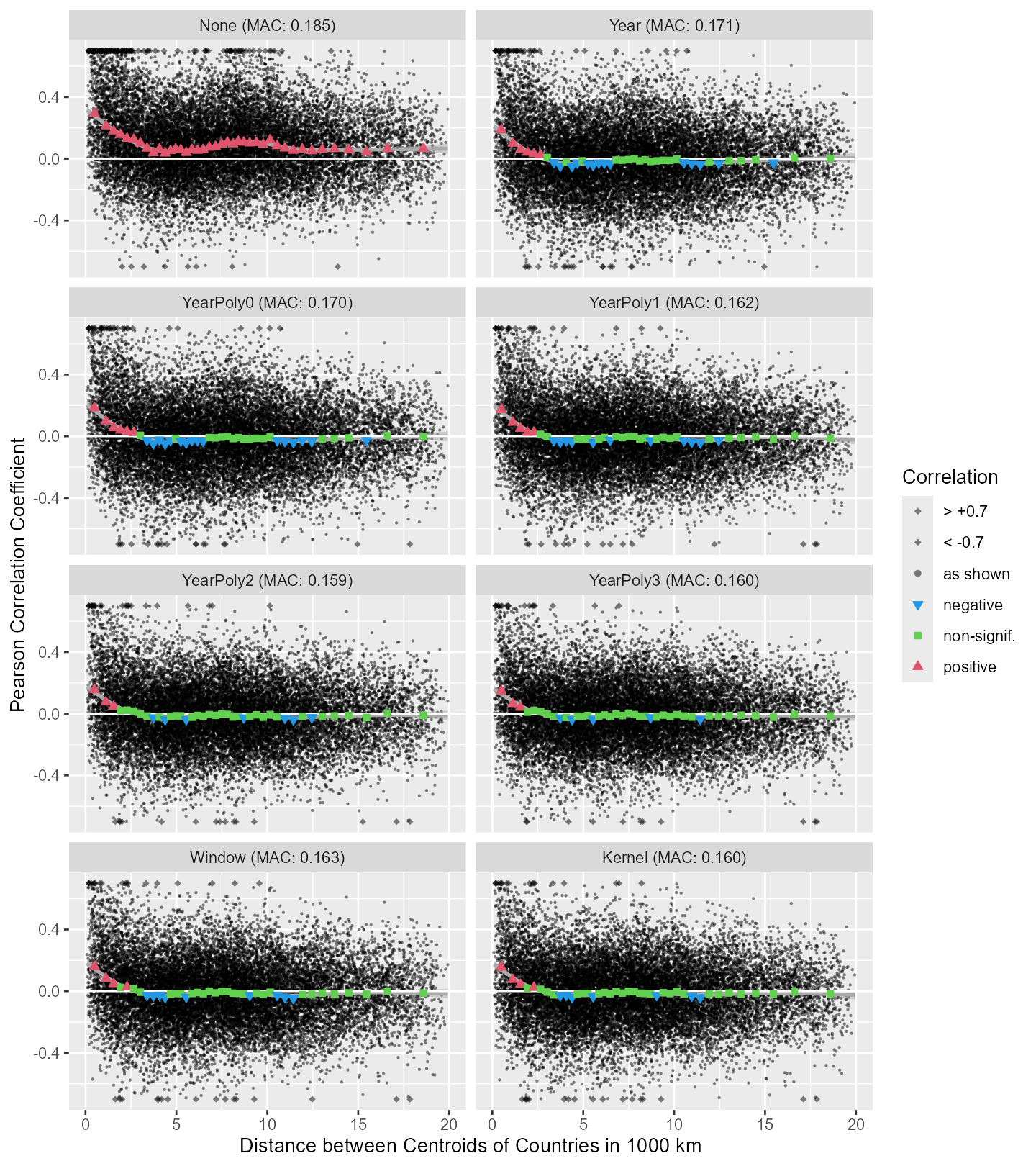}
	\caption{Effect of different control variable designs on spatial correlation of residual economic growth after controlling. Each black dot is an empirical correlation value between two different countries calculated from the ISIMIP economic data. On the horizontal axis, the distance between the centroids of the two countries is shown. The gray curve shows a Loess smoother of the correlations. For 40 bins of spatial distances with roughly equal number of pairs of countries, we test whether the correlations values for pairs of countries are significantly different from zero via a conservative permutation test, see \supplementref{section S4}. Furthermore, we apply the Benjamini-Hochberg correction (with false discovery rate level of $0.05$) to account for the multiple testing issue. Blue and red triangles show correlations significantly different from zero, whereas green squares indicate that non-zero correlation cannot be detected. MAC is the mean absolute correlation. Without controls (\textit{None}), we observe a high positive correlation for any spatial distance. As soon as year fixed effects are applied, only low distance positive correlations remain, but some negative correlations for larger distances get newly introduced. Adding polynomial or local time trends to the controls reduces the negative correlations. From this perspective, local time trends or polynomial time trends of degree at least $1$ should be preferred.}
	\label{fig:fe:corrspat}
\end{figure}

\Cref{fig:fe:corrtemp,fig:fe:corrspat} contrast the temporal and spatial correlations, respectively, resulting from different control variable designs. The local linear kernel approach yields the least systematic temporal and spatial correlations and does not introduce spurious correlation artifacts over extended time periods. Furthermore, using the automatic bandwidth choice, we remove further control design decisions ensuring objectivity of results. 

\subsection{Correlations and Standard Errors}

Our investigation of correlations focuses primarily on the economic growth variable and their residual values after controlling for global shocks and regional time trends. This approach is justified by the fact that, after controlling, the estimated residuals are nearly identical to the residual of a full regression model with climate variables. The reason is, that in our application no climate variable explains a large portion of the variation in economic growth. This also means that a preliminary correlation analysis for the climate variables is not crucial. Nonetheless, should we identify a highly predictive (climate) variable, it would be appropriate to redo the analysis of correlations then based on the residuals from the final regression model.

\subsubsection{Different Standard Error Designs}

After controlling, some correlation in the error term typically remains. In order to ensure valid inference, we must choose a standard error estimator that properly accounts for the remaining correlations. In our work, we have considered several approaches, each of which deals with spatial and/or temporal dependence in distinct ways. In what follows we briefly summarize these methods and their key properties.

\paragraph{HAC (Newey--West) Standard Errors.}
The Newey--West estimator is specifically designed to address auto-correlation in time series data. It applies a kernel (typically the Bartlett kernel) that weights observations based on their temporal distance, thereby correcting for serial correlation and heteroskedasticity. However, Newey--West standard errors do not address cross-sectional (spatial) dependence.

\paragraph{Clustered Standard Errors by Year.}
Clustering standard errors by year allows for arbitrary correlation among cross-sectional units (e.g., countries) within the same time period. This approach assumes that, within each year, the error terms may be correlated in an unrestricted manner across units, but that there is no serial (temporal) correlation across different years.

\paragraph{Clustered Standard Errors by Country.}
In contrast, clustering by country accommodates arbitrary temporal correlation within each country. Here, it is assumed that errors within a given country over time may be arbitrarily correlated, while errors across different countries are treated as independent. In contrast to estimators that deal with temporal auto-correlation where the correlation is determined by the lag (difference between two years), here any two years can be arbitrarily correlated.

\paragraph{Two-Way Clustered Standard Errors by Country and by Year.}
Multi-way clustered standard errors \citep{Cameron01042011,thompson11} require uncorrelated error terms only for observations that do not share a common cluster in any clustering dimension. In our case, we require an observation from country $\reg$ and year $\tim$ to be uncorrelated with an observation from country $\reg\pr$ and year $\tim\pr$ for $\reg \neq \reg\pr$ and $\tim\neq\tim\pr$. But we account for correlations between observations at $(\reg, \tim)$ and $(\reg, \tim\pr)$ as well as for correlations between $(\reg, \tim)$ and $(\reg\pr, \tim)$. To have a reliable estimate of two-way clustered standard errors, we need to have enough clusters for each clustering dimension (enough different year, and enough different countries).

\paragraph{Conley Standard Errors.}
While Newey--West account for temporal correlation based on temporal distance, Conley standard errors account for spatial correlation based on spatial distance. A kernel is applied that downweights correlations as the distance increases. The distance is typically taken as the physical distance, but if other notions of distance exist, they can also be used. Some implementations can additionally deal with temporal auto-correlation (in R, the package \texttt{conleyreg}).

\paragraph{Driscoll--Kraay Standard Errors.}
Driscoll--Kraay standard errors extend the HAC idea by first averaging the estimating equations over the cross-sectional units and then applying a Newey--West type correction. This method is robust to both arbitrary cross-sectional (spatial) correlation (not necessarily based on distance) and temporal auto-correlation (based on temporal distance). A potential drawback is that this estimator is more data hungry (i.e., it requires a sufficiently long time dimension) and tends to be less precise when the number of time periods is small.

\subsubsection{Correlations in the Economics Growth Rate}
We emphasizes that spatial correlation in economic data is often not driven solely by the physical distance between countries. While neighboring countries frequently exhibit strong trade links (which is an argument in favor of distance based spatial correlation), being members of the same economic union or sharing similar economic policies can induce strong correlations even if the countries are not immediate neighbors or geographically close.

To demonstrate this point, we compare correlations between countries of the European Union (EU) with correlations of countries that are geographically close. We use the countries of the EU that were members in the year 1996 (EU membership does not change thereafter until 2004) so that the effect of being part of the EU shows in the observation time series (1967 to 2018). The mean correlation between non-neighboring EU countries is \key{corr}{meanEuCoreNonNeighbors} whereas neighboring countries globally have an average correlation of \key{corr}{meanNeighbors}. This difference is highly significant ($p < 10^{-5}$ according to a bootstrap test). Similarly, the mean correlation between EU countries with a centroid distance larger than the median EU centroid distance (\key{corr}{mediDistEuCore}) is \key{corr}{meanEuCoreDistant}, which is significantly ($p < 10^{-5}$ according to a bootstrap test) larger than \key{corr}{meanAllClose}, the global mean correlation of countries with less centroid distance than the aforementioned threshold. Thus, being part of a common economic zone such as the EU is more relevant than geographical proximity.

Furthermore, our residual analysis reveals that there are a some instances of long-distance (negative) spatial correlations (\cref{fig:fe:corrspat}), while long temporal correlations are absent (\cref{fig:fe:corrtemp}) when using \textit{Kernel} controls. In view of these features, we advocate for the use of Driscoll--Kraay standard errors. Despite their higher data requirements, they offer a robust solution that simultaneously addresses both arbitrary spatial correlation and temporal auto-correlation, without suffering from the systematic bias observed in alternative estimators.

\subsubsection{Small Sample Correction}
To account for the degrees of freedom already removed in the controlled data set, the small sample correction in the calculation of standard errors has to be adjusted appropriately. In the case of local linear time trends, we use the effective number of parameters, i.e., the trace of the smoother matrix $S$, which fulfills $\mo{\hat Y} = S \mo{Y}$, where $\mo{\hat Y}$ is the temporally smoothed economic data, so that the controlled data is $\mo{\tilde Y} = \mo{Y} - \mo{\hat Y}$.

%% file: sec_selection.tex
\section{Step 3: Model Selection via Out-of-Sample Test}\label{sec:modelselection}
Traditionally, studies have relied on annual averages of temperature (\vn{tas}) and precipitation (\vn{pr}) to predict GDP growth. However, such models may be overly simplistic. More recent research has introduced complex variables—often derived from \vn{tas} and \vn{pr}—such as measures of extreme rainfall or temperature fluctuations. In addition, studies examining specific channels through which weather impacts the economy often incorporate variables like the number of people affected by tropical cyclones.

When accounting for all plausible combinations and transformations of these variables, the number of potential predictors grows dramatically. Selecting a model based solely on its fit, without acknowledging the many alternatives considered and discarded, risks $p$-hacking and fosters unwarranted confidence in the results. This underscores the necessity of a formal, transparent variable selection procedure.

In the analysis that follows, we implement a suite of formal model selection procedures. Crucially, model performance is assessed using out-of-sample metrics, rather than in-sample fit, which can be misleading due to reliance on assumptions that may not hold in practice. This approach provides a more robust evaluation of predictive accuracy.

Variable selection and testing are carried out on the controlled dataset, allowing us to isolate the effect of the predictors from that of the control variables. To assess the robustness of the results, we repeat the analysis across multiple datasets, each defined by a different control specification. However, it is important to emphasize that test-set performance cannot be directly compared across different control variants, as each defines a distinct prediction objective.
\subsection{Methods}
The controlled data is split into training (pre-2007) and testing (2007 onward) subsets, yielding 41 years of training and 11 years of testing data. To assess robustness, we also evaluate alternative splits using 1997 and 2002 as threshold years. We avoid shorter test periods, as they reduce the reliability of evaluation statistics. The split at a threshold year maintains spatial dependencies and reduces the influence of temporal auto-correlation. Furthermore, it exactly mimics the main use case of the model results in which economic damages are projected into the future.

Most selection methods involve tuning hyperparameters, which we optimize via five-fold cross-validation on the training set. Folds are composed of consecutive years, ensuring that all data from a given year remains in the same fold. This design again respects both spatial correlation and reduces the influence of temporal auto-correlation.

Once optimal hyperparameters are identified, each model is retrained on the full training set and used to predict the (controlled) economic response in the test set.  We assess the predictive significance of each model by benchmarking its performance against a random prediction independent of any predictor variables and modeled by an isotropic Gaussian random vector with optimally chosen variance. See \supplementref{section S6} for details.

To assess the stability of predictor selection, we employ a block bootstrap approach, resampling the training data 1000 times. Each resample includes a number of years equal to the original training period, drawn with replacement. When a year is drawn multiple times, its data appears multiple times in the resampled dataset. For each bootstrapped sample, we reapply the selection procedure (but not the hyperparameter tuning) and track how often the originally selected predictors are re-selected. This \textit{re-selection rate} serves as an indicator of the robustness of the variable selection process.

Note that all of these calculations happen on the controlled data. This is necessary, as we cannot estimate the controls of the test data using the training data, and all analysis has to be carried out orthogonality to the controls. We expand on this issue and compare our framework with that of \citet{newell_gdp-temperature_2021} in \supplementref{section S7}.
\subsubsection{Information Criteria}
A common way of model selection is using information criteria such as the Akaike information criterion (AIC) and the Bayesian information criterion (BIC). Unfortunately, these are not applicable in our setting as they require independent and identically distributed variables. We instead have a complex dependence structure with temporal and spatial correlation and possible heteroskedasticity. Extending the information criteria to such dependence structures is highly non-trivial. Therefore, we do not use these criteria in our experiments.
\subsubsection{Subset, Pair, and Forward Selection}
In \emph{Subset Selection}, we fix the number of predictors $0 \leq k \leq p$. For each subset of $k$ predictors, we fit the model using ordinary least squares (OLS) and compute the residual sum of squares (RSS), selecting the subset that yields the lowest RSS. The value of $k$ is a hyperparameter to be chosen via cross-validation. However, because there are ${p \choose k}$ possible subsets of size $k$ from $p$ predictors, this approach is computationally infeasible for $k$ larger than $2$ or $3$: For our case with $p = 708$, the number of subsets for $k = 2, 3, 4$ is approximately $2.5 \times 10^5$, $5.9 \times 10^7$, and $1.0 \times 10^{10}$, respectively.

To address this, we instead employ a more computationally efficient variant known as \emph{Forward Selection} \citep{Hastie2009}. This method builds the model sequentially by adding, at each step, the single predictor that most reduces the RSS. While this approach drastically reduces computation time, it has a key limitation: It may fail to select predictor combinations that are jointly informative but not individually strong. For instance, a variable like $\vn{tas}$ and its squared term $\vn{tas}^2$ might only be useful when included together, and Forward Selection might overlook them individually. In contrast, Subset Selection could capture such combinations.

To assess the impact of this limitation, we also implement \emph{Pair Selection}, i.e., Subset Selection restricted to $k = 2$, to evaluate whether jointly informative pairs are being missed by Forward Selection.
\subsubsection{LASSO, Ridge, and Elastic Net}
The \emph{LASSO} and \emph{Ridge Regression} penalize the absolute values of the regression coefficients so that small absolute values are preferred. For a hyperparameter $\lambda\geq0$, the respective coefficient estimates can be written as
\begin{align}
	\hat\beta_{\ms{LASSO}} &=\argmin_{\beta} \brOf{\frac1n\Vert \tilde{\mo Y} - \tilde{\mo Z}\beta \Vert_2^2 + \lambda \Vert\beta\Vert_1}\,,\\
    \hat\beta_{\ms{Ridge}} &=\argmin_{\beta} \brOf{\frac1n\Vert  \tilde{\mo Y} - \tilde{\mo Z}\beta \Vert_2^2 + \frac\lambda2\Vert\beta\Vert_2^2}\,.
\end{align}
The penalties allow these methods to be applied in settings with a large number of coefficients compared to the sample size. Meaningful estimates are possible even in settings with more coefficients than observations. The different penalties between LASSO ($L^1$) and Ridge ($L^2$) yield different coefficient estimates: LASSO prefers to set many coefficients to exactly $0$ (thus implying a selection of variables), whereas Ridge typically does not estimate any coefficient to be exactly $0$ but all coefficients will be shrunken sufficiently close to $0$. Note that both methods tend to estimate coefficients with lower absolute values than OLS, i.e., they imply a form of shrinkage. The \emph{Elastic Net} combines both LASSO and Ridge Regression:
\begin{equation}
    \hat\beta_{\ms{ElasticNet}} =\argmin_{\beta} \brOf{\frac1n\Vert \tilde{\mo Y} - \tilde{\mo Z} \Vert_2^2 + \lambda \br{\frac{1-\alpha}2\Vert\beta\Vert_2^2 + \alpha\Vert\beta\Vert_1}}\,.
\end{equation}
The penalty parameter $\lambda\geq0$ and the trade-off parameter $\alpha\in\{0.1, 0.2, \dots, 1\}$ for ElasticNet are chosen using $5$-fold cross validation on the training set with the folds consisting of all data from mutually exclusive consecutive years. Ensuring $\alpha>0$ promotes ElasticNet to set some coefficients to exactly $0$ as happens in LASSO implying variable selection.
\subsubsection{Fordge}
As pair and forward selection yield much worse results than the shrinkage methods LASSO, Ridge, and Elastic Net, we hypothesize that the missing shrinkage is the problem. Thus, we additionally test the usage of Ridge regression on the variables chosen by forward selection. We call this procedure \emph{Fordge} (a portmanteau of forward and ridge).

%% file: sec_results.tex
\section{Results}
Here we present the results of all six different model selection methods in different settings. Detailed results can be found in \supplementref{section S10}.
\subsection{Test Error}

\paragraph{All Experiments.}
First, we consider the mean squared error (MSE) on the test set and its significance for our six methods: Ridge Regression, LASSO, Elastic Net, Pair Selection, Forward Selection and Fordge. Apart from different methods, we also compare raw vs cleaned data, ISIMIP vs World Bank GDP, $8$ different control designs, level vs growth regression, and three different options for the threshold year that splits our data into train and test set. In total, this makes 1152 experiments. 

For binary attributes the results are summarized in \cref{tbl:summary:vs}.
Generally, the cleaned data is clearly easier to predict (in 197 cases with 337 draws and 42 opposite cases). This is to be expected as outliers are difficult or even impossible to forecast.
The World Bank data seems somewhat more predictable than the ISIMIP data (in 136 cases with 369 draws and 71 opposite cases), but the difference is not large.
Level and growth regression have similar predictability (359 draws, 93 cases where level is more predictable, 124 growth).

Predictive power slightly improves when increasing the train-test split year from 1997 to 2007: The mean test error ranks among the three threshold choices are 2.07 (split in 1997), 1.97 (split in 2002), 1.95 (split in 2007), respectively. Furthermore, the number of experiments with any predictive power (test error lower than for the trivial model without predictors) increases with later split year. Out of 384 experiments nonzero predictive power is shown in 82 (split in 1997), 96 (split in 2002), and 112 (split in 2007) experiments, respectively. Both results seem to be a consequence of better estimates due to more training data for later split years.

The average predictability of data with different control does not differ a lot with \emph{Window} having the highest predictability. Note that high predictability is not necessarily a quality criterion for the control design.

The test error results for different model selection methods are collected in \supplementref{section S8}, which we further summarize in \cref{tbl:summary:methods}.
Ridge regression shows the highest predictive power followed by Elastic Net and LASSO, while Pair Selection, Fordge and Forward Selection show worse performances. 

\begin{table}
	\input{tbl/09_together_vsSummary.tex}
	\caption{Summary of experiments along different attributes. We evaluate the binary attributes \textit{preprocessing}, \textit{regression design}, and \textit{GDP data source} by comparing the $p$-value on the test set for both instances of the attribute and counting how often each instance of the respective attribute has a lower value than the other instance. The high number of draws (both versions perform equally) is due to the $p$-value being set to $1$ if the method's test error is larger than the variance of the test data, i.e., the test error of the trivial constant prediction.}
	\label{tbl:summary:vs}
\end{table}

\begin{table}
	\input{tbl/09_together_pValCount_methods.tex}
	\caption{Summary of performance of different methods according to their test error rank (1 to 6) averaged over different experimental settings. All comparisons are made from all 1152 experiments. For the main comparisons we restrict to cleaned data, kernel controls, and 2007 as the split year and vary only between level and growth regression designs and ISIMIP and World Bank as GDP data sources.}
	\label{tbl:summary:methods}
\end{table}

\paragraph{Main Experiments.}
To ensure that this evaluation is not diluted by averaging over different controls and mixing cleaned and uncleaned data, we next focus only on our preferred settings with cleaned data, Kernel controls, and train-test split in 2007---we deem those to be the most relevant and reliable. This leaves 24 experiments (6 methods, ISIMIP vs World Bank, and level vs growth regression). 

For the two binary attributes (GDP data source and regression design), the results are summarized in \cref{tbl:summary:vs}.
We find the ISIMIP data to be slightly more predictable compared to the World Bank data (in 5 cases with 5 draws and 2 opposite cases). This is opposite to the evaluation of all experiments and indicates that the differences are not a strong structural property of the datasets but rather random fluctuations.
In 7 out of 12 cases (with 3 draws), level regression models have higher predictive power than growth regression models. This pattern does not show in the common evaluation of all experiments. So, it is unclear how far this statement generalizes.

The three penalized regression methods---Ridge, LASSO, and Elastic Net---again outperform the other approaches (see \cref{tbl:summary:methods}). Therefore, we conclude that Ridge, LASSO, and Elastic Net should be preferred. Among them, Ridge consistently achieves the best performance. Nonetheless, one might still favor LASSO or Elastic Net due to their greater interpretability, as they perform explicit variable selection---unlike Ridge Regression.

\paragraph{Best Test Error.}
The most predictive setup of the main experiments (level regression, Elastic Net, ISIMIP data) is only able to explain $0.19\%$ of the variance in the test data. This is equivalent to the amount achieved in $2\%$ of cases using a prediction vector independent of predictor variables with a randomly chosen direction and optimal length (i.e. $p = 0.02$). Consequently, despite the inclusion of $708$ variables (in both original and differenced forms), a significant proportion of which have been previously associated with GDP in other studies, there remains a small possibility that the predictor dataset contains no substantial information regarding GDP.

\subsection{Selected Variables}
To evaluate which among the 708 variables in our regression are most important, we focus on the main experiments (cleaned data, Kernel controls) considering level and growth regression and ISIMIP as well as World Bank data sources for GDP. Furthermore, as we here do not need a test set for this evaluation, we use all data for training, i.e., until 2018, to get the most accurate results. We compare the four out of six model selection methods that explicitly select variables (Forward, Pair, LASSO, Elastic). The results are summarized in \supplementref{section S9}.

There is only a single variable that gets chosen systematically (except by Forward on growth regression, which does not select any variable): The square of the maximal specific humidity of the previous year, $\mathsf{huss}_{\mathsf{max},-1}^{2}$  (or $\Delta\mathsf{huss}_{\mathsf{max},-1}^{2}$ in the level regression case). This variable also has the highest bootstrap re-selection rate among all chosen variables in all settings. 
Reasons for the relatively high predictive power could be the correlation of high specific humidity to  heat stress, storms, and flooding, as well as correlation of low specific humidity with droughts, water shortages, wild fires, respiratory issues and airborne pollutants. 

No other variable is chosen robustly. In particular, classical variables such as $\vn{tas}$, $\vn{tas}^2$, $\vn{pr}$, $\vn{pr}^2$ are never chosen by any algorithm but some transformations of \vn{tasmin} and \vn{tasmax} do show up for some settings without strong consistency.

\subsection{Shrinkage}

LASSO, Ridge Regression, and Elastic Net typically perform better than the other three methods and Fordge is often slightly better than Forward Selection. This suggests that penalized regression in general has an advantage in our setting. To investigate this hypothesis further, we select specific variables either classical $\vn{tas}$, $\vn{tas}^2$, $\vn{pr}$, $\vn{pr}^2$ or preferred by our variables selection methods ($\vn{huss}_{\mathsf{max},-1}^{2}$, $\vn{sfcwind}_{\mathsf{max}}^{2}$) and use OLS as well as LASSO and Ridge fits for these variables. \Cref{tbl:shrinkage} shows the results. In all cases the shrinkage of coefficients implied by the penalty of LASSO and Ridge helps to achieve lower test errors.

\begin{table}
	\input{tbl/10_shrinkage.tex}
	\caption{For a selected number of different sets of predictor variables, we calculate the OLS, LASSO and Ridge fit, using 5-fold cross validation to determine the penalty parameter for the latter two. We use the cleaned ISIMIP data with Kernel controls and split it into train and test data in 2007. The columns \textsf{MSE\%} show the respective change in percent in mean squared error on the test set compared to the trivial null-model. The columns \textsf{vs OLS} show the change of \textsf{MSE\%} from OLS to LASSO and Ridge, respectively.}
	\label{tbl:shrinkage}
\end{table}

\subsection{Stationarity}

In our regression analysis, we model the proportional influence of shocks in the climate variables on the economic growth rate to be constant over time and space (temporal and spatial stationarity). This, of course, is an approximation. Even if this does not perfectly reflect the real world, such a model may still be useful, as it allows us to estimate an average effect. But this approximative nature of our analysis has to be kept in mind when interpreting the results or using them for projections into the future.

\begin{figure}
	\begin{center}
        \includegraphics[width=\textwidth]{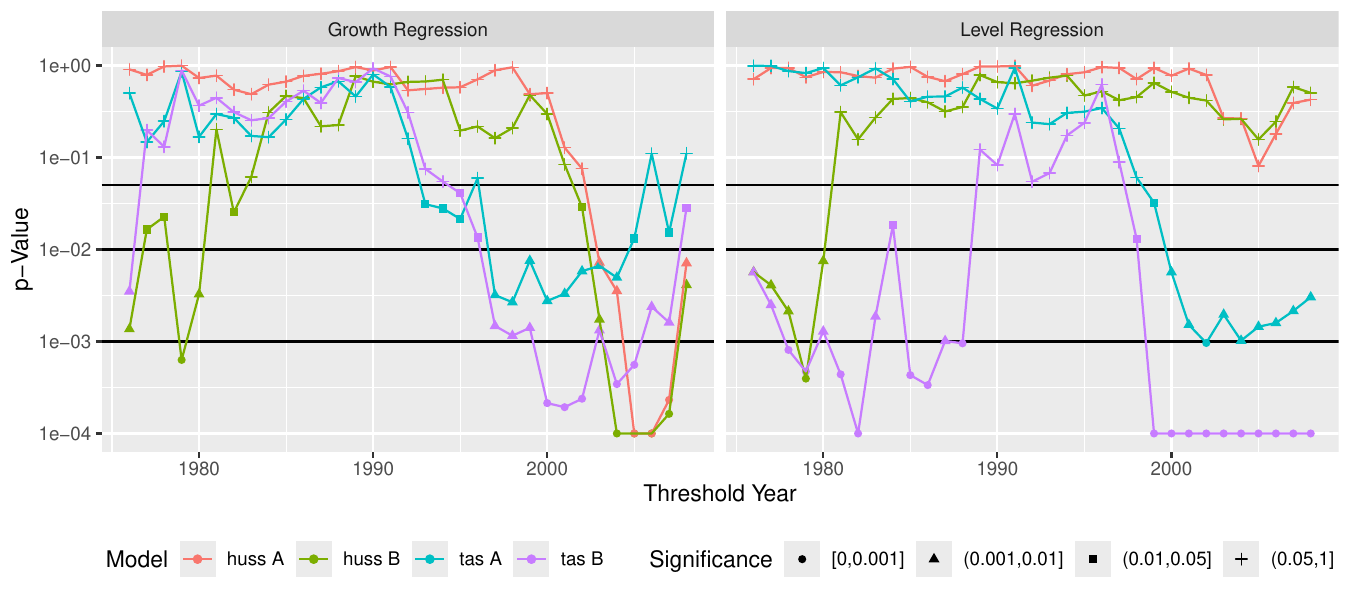}
    \end{center}
	\caption{$p$-values of $F$-test for stationarity of effects. We split the data (Kernel controls, cleaned ISIMIP economic data) into two parts: up until the year indicated on the horizontal axis and afterward. Then, for both parts, we fit the OLS models indicated by the color of the displayed curves, see \cref{tbl:shrinkage}. We test the Null-hypothesis of equal coefficient vectors for both parts $\beta_{\ms{before}} = \beta_{\ms{after}}$ using an $F$-test for linear hypothesis and the Driscoll-Kraay covariance matrix with $8$ lag years. The $p$-value of the test is shown via the vertical axis, which is on a log-scale and cut at $10^{-4}$. Using different symbols, we indicate different significance levels (their threshold is displayed by black horizontal lines). We only show threshold years from 1977 to 2007, even though the data ranges from 1967 to 2018, to ensure that there is enough data on each side of the threshold year to make the results reliable.}
	\label{fig:discuss:stationarity:thershold}
\end{figure}

To ensure validity of future projections based on such regression results, we do require temporal stationarity: The weather--economy relationship should be stable over the observation time to reasonably assume that this relationship also exists in the future. We investigate this property for models involving the classical temperature parabola \citep{burke_global_2015} and models involving the humidity-variable favored by our variable selection analysis. With an F-test, we check whether the data before a threshold year exhibits a similar regression coefficient than the data thereafter, see \cref{fig:discuss:stationarity:thershold}. Most models show significantly different effects in the first time period compared to the later time period, if the threshold year is chosen between 2000 and 2010 indicating a change point within these years. Only the pure specific humidity model (huss A) for level regression is stable over the whole training period. Adding maximal wind surface speeds to the level regression humidity model (huss B), seems to exhibit a change point before  1980 but stability thereafter. These two models show the greatest temporal stationarity, which is consistent with their low test errors in \cref{tbl:shrinkage}. 

Investigating the change of coefficients in the temperature parabola models further reveals that the quadratic influence of temperature on the economic growth seems to strongly reduce after 1992, see \supplementref{Figure S2}. Thus, when measuring predictive power with a test set of some years after 1993, the quadratic relationship learned from previous years is not predictive. This instability is reason for concern when projecting economic influence of temperature to the future. Coefficient values of the level regression humidity model (huss A), seem slightly more stable, in particular when estimated over time periods of at least 15 years.

Spatial stationarity appears less critical than temporal stationarity, as we are typically projecting outcomes for existing countries rather than entirely unseen ones. If the goal is merely to estimate the average effect of weather on the economy, spatial stationarity is not required. However, to make country-specific projections, spatial stationarity either must be demonstrated or it must be clearly stated that the projections do not reflect country-specific economic structures. Instead, they represent average outcomes for a hypothetical country with similar climatic conditions but globally averaged economic characteristics---ignoring factors such as the level of industrialization or the share of agriculture in economic output of the specific country.

It is also important to note that calculating global GDP impacts represents a form of averaging over countries, though with a different weighting scheme than that used in model estimation. While the model gives equal weight to each country---regardless of economic size, population, or geography---global damage estimates are typically dominated by the losses of wealthier countries due to their larger contributions to global GDP.

%% file: sec_discussion.tex
\section{Discussion}

This paper advances climate econometrics by addressing key methodological shortcomings in existing research and proposing data-driven solutions. Our enhanced empirical strategies challenge earlier findings and underscore the need for more robust modeling practices. To improve future research, we provide methodological recommendations for future climate econometric research.

\textbf{Outlier treatment and data integrity.}
We document the substantial influence of outliers on empirical results, present procedures for their identification and treatment, and provide a cleaned dataset to facilitate more reliable analyses. Only few previous works \citep{pretis_uncertain_2018} treat outliers at all. 

\textbf{Flexible time-trend controls and correlations.}
We replace arbitrary, parametric time-trend specifications with a nonparametric, fully automatic procedure. This reduces practitioner bias and more effectively captures complex temporal dynamics in panel data. Furthermore, we show the influence of different control designs revealing undesirable long term temporal correlations introduced by previously common polynomial time trends.

\textbf{High-dimensional variable selection with out-of-sample validation.}
To navigate over 700 candidate predictors, we employ formal variable-selection methods—most notably LASSO, Ridge, and Elastic Net—which guard against overfitting, $p$-hacking, and cherry picking. Crucially, we benchmark all models via out-of-sample testing, providing a realistic assessment of predictive power under unknown dependency structures and avoiding untested assumptions inherent in classical inference methods.

\textbf{Empirical findings.}
Strikingly, almost none of the tested variables—including those emphasized in prior studies—retain meaningful predictive power out-of-sample. A key reason appears to be the violation of temporal stationarity, an assumption often made but rarely tested. Among the penalized methods, a humidity-related variable consistently emerges as the strongest predictor of economic damage. Investigating this variable—and its interactions with temperature—from both statistical and mechanistic perspectives is a promising avenue for future work.

\textbf{Implications.}
By integrating outlier robustness, adaptive trend controls, rigorous variable selection, and out-of-sample validation, our approach reveals the fragility of many conventional findings and calls for a paradigm shift toward methodologies that prioritize predictive accuracy and generalizability in climate econometrics.

\textbf{Recommendations.}
From our finding, we derive following recommendations for future work in climate econometrics to improve robustness and avoid wrong conclusions.
\begin{enumerate}[label=\arabic*.]
    \item Check the data for outliers and analyze their influence.
    \item Check temporal and spatial correlations in controlled data to be able to choose a suitable standard error and cross-valiation design, and to be able to compare different control designs.
    \item 
    Use a formal model selection procedure to avoid a subjective model choice or cherry picking (e.g., use LASSO, Ridge Regression, or Elastic Net).
    \item 
    Use out-of-sample testing that is robust against potential dependence structures in the data to verify that the chosen model generalizes.
    \item 
    Check temporal stationarity if results are used for future projections.
    \item 
    Check spatial stationarity if results are interpreted as country specific or interpret results as global averages only. 
    \item 
    Try shrinking estimated coefficients, e.g., using penalized regression methods (such as LASSO, Ridge Regression, or Elastic Net) to improve predictive power and mitigate against problems with data dependency and a lack of stationarity.
\end{enumerate}

%% file: app_details.tex
\clearpage
\section{Data Summary Statistics}\label{sec:data}

In \cref{tbl:data:growth,tbl:data:one:o,tbl:data:two:o,tbl:data:one:d,tbl:data:two:d}, we report summary statistics for all variables used in this study (excluding lagged variables), aggregated across all countries and years. For each variable, we provide the mean, standard deviation (StdDev), skewness, and kurtosis, as well as the following quantiles: minimum (Min), first quartile (Q1), median, third quartile (Q3), and maximum (Max). Note that these statistics are computed directly from the raw input data, whereas the coefficient magnitudes discussed in the main text may refer to scaled versions of these variables.

\begin{table}[h!]
    \input{tbl/04_dataStats_log2growth.tex}
    \caption{See \cref{sec:data} for a description.}\label{tbl:data:growth}
\end{table}
\begin{table}
    \input{tbl/04_dataStats_1_o.tex}
    \caption{See \cref{sec:data} for a description.}\label{tbl:data:one:o}
\end{table}
\begin{table}
    \input{tbl/04_dataStats_2_o.tex}
    \caption{See \cref{sec:data} for a description.}\label{tbl:data:two:o}
\end{table}
\begin{table}
    \input{tbl/04_dataStats_1_d.tex}
    \caption{See \cref{sec:data} for a description.}\label{tbl:data:one:d}
\end{table}
\begin{table}
    \input{tbl/04_dataStats_2_d.tex}
    \caption{See \cref{sec:data} for a description.}\label{tbl:data:two:d}
\end{table}
\clearpage
\section{Details on Examples}\label{sec:details}

\begin{table}[h!]
	\input{tbl/01_DOSE_Spain.tex}
	\caption{Economic Output in USD 2015 (column \texttt{grp\_pc\_lcu2015\_usd} in the dataset) for three Spanish regions (\cref{tbl:dose:spain:meta}) in Version 2 and 4 (there is no Version 3) of the DOSE dataset \citep{dose}. Corrections were made in Version 4 that alter the economic output almost by a factor of 10. These corrections also remove an unrealistic changes of economic output between 1994 and 1995 of $+722\%$ (ESP.9\_1) and $-89\%$ (ESP.8\_1) in Version 2.}\label{tbl:dose:spain}
\end{table}
\begin{table}[h!]
	\input{tbl/01_DOSE_Spain_Meta.tex}

	\caption{The identifiers for the Spanish regions shown in \cref{tbl:dose:spain}.}\label{tbl:dose:spain:meta}
\end{table}
\begin{table}[h!]
	\input{tbl/02_ISIMIP_IRQ_1990-2005.tex}
	\caption{GDP per capita in USD, rounded to 3 significant digits, for Iraq between 1990 and 2005 according to the ISIMIP data.}\label{tbl:isimip:iraq}
\end{table}
\begin{figure}[h!]
    \includegraphics[width=\textwidth]{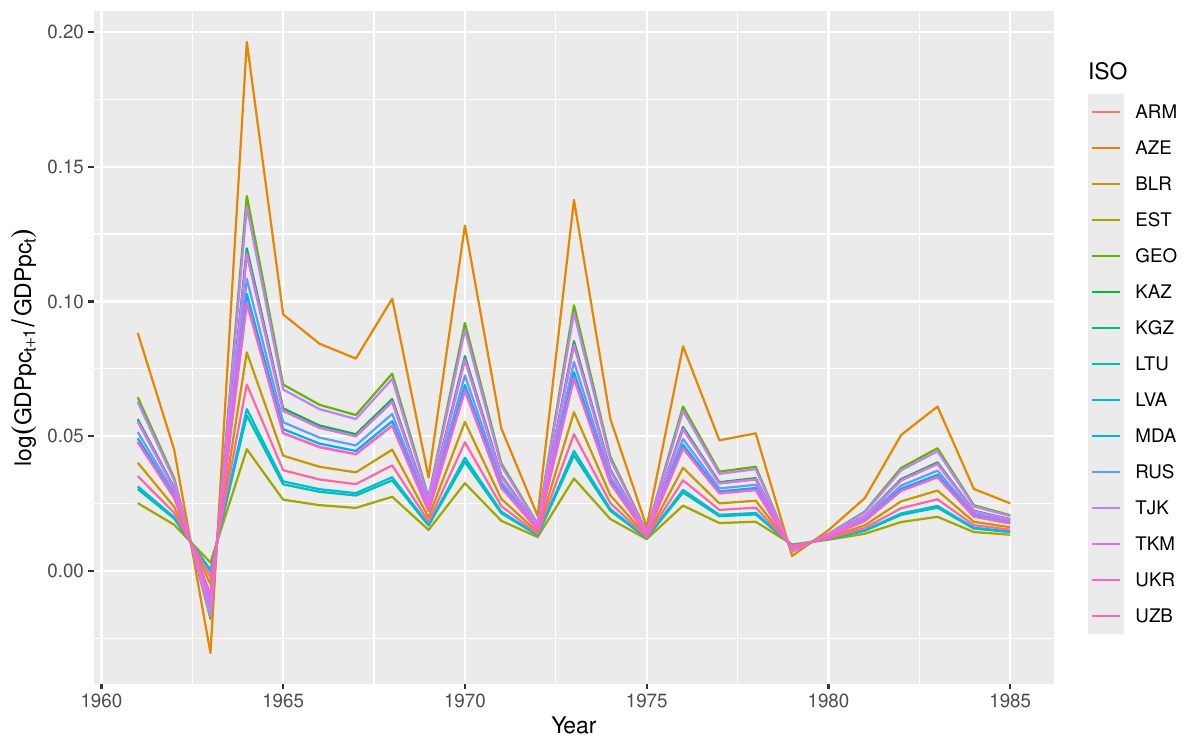}
    \caption{Imputed data for economic growth of post-Soviet states in the ISIMIP dataset before 1986. See \cref{tbl:isimip:meta:a,tbl:isimip:meta:b,tbl:isimip:meta:c} for a translation of country codes to country names.}\label{fig:isimip:soviet}
\end{figure}
\begin{table}[h!]
	\input{tbl/01_iso_1.tex}
	\caption{The country codes used in the ISIMIP dataset.}\label{tbl:isimip:meta:a}
\end{table}
\begin{table}[h!]
	\input{tbl/01_iso_2.tex}
	\caption{The country codes used in the ISIMIP dataset.}\label{tbl:isimip:meta:b}
\end{table}
\begin{table}[h!]
	\input{tbl/01_iso_3.tex}
	\caption{The country codes used in the ISIMIP dataset.}\label{tbl:isimip:meta:c}
\end{table}
\begin{table}[h!]
	\input{tbl/02_clusters_min10.tex}
	\caption{Clusters of countries with highly pair-wise correlated economic time series in the ISIMIP dataset. The target variable is the economic growth given by $Y_{\reg,t} = \log(\ms{GDPpc}_{\reg,t}/\ms{GDPpc}_{\reg,t-1})$. The column \emph{avg. cor} shows the correlation averaged over all pairs of different countries in the cluster. See \cref{tbl:isimip:meta:a,tbl:isimip:meta:b,tbl:isimip:meta:c} for a translation of country codes (\emph{regionIds in Cluster}) to country names.}\label{tbl:isimip:clusters}
\end{table}
\begin{table}[h!]
	\input{tbl/02_const.tex}
	\caption{Countries and  year ranges (of at least two years) without economic change, $\absof{Y_{\reg,t}} \leq 2\cdot 10^{-13}$, where $Y_{\reg,t} = \log(\ms{GDPpc}_{\reg,t}/\ms{GDPpc}_{\reg,t-1})$. See \cref{tbl:isimip:meta:a,tbl:isimip:meta:b,tbl:isimip:meta:c} for a translation of country codes (\emph{regionIds}) to country names.}\label{tbl:isimip:const}
\end{table}
\clearpage
\section{Robust Regression}\label{sec:robust}
We attempt to perform a robust regression on the model equation (6) of the main article using the raw ISIMIP data. The ordinary least squares (OLS) estimates can be computed in R using the \texttt{lm()} function. Several R packages offer functionality for robust regression. However, these packages often do not support singular fits, and a straightforward inclusion of fixed effects can result in singular model matrices. To address this, it is necessary to remove an appropriate number of columns from the model matrix before applying these functions. This adjustment can be achieved using a QR decomposition, as demonstrated below:
\begin{verbatim}
    # frml contains the regression formula; data is a dataframe
    X <- model.matrix(frml, data = data)
    qr_decomp <- qr(X)
    X_full_rank <- X[, qr_decomp$pivot[1:qr_decomp$rank]]
\end{verbatim}
Note that this does not change the results:
\begin{verbatim}
    fit1 <- lm(y ~ X - 1)
    fit2 <- lm(y ~ X_full_rank - 1) # yields the same coefficients as fit1
\end{verbatim}
With the full rank model matrix, a robust regression fit can be obtained via 
\begin{verbatim}
    MASS::rlm(y ~ X_full_rank - 1, method = "M")
\end{verbatim}
But this took more than 10 times as long as the OLS fit. Alternative robust regression methods such as
\begin{verbatim}
    MASS::rlm(y ~ X_full_rank - 1, method = "MM")
\end{verbatim}
and
\begin{verbatim}
    robustbase::lmrob(y ~ X_full_rank - 1, n.group = ncol(X_full_rank)*2)
\end{verbatim}
failed due to numerical problems.

In the results of the robust regression, certain data points are weighted down significantly. For example, the 1988 observation for Nicaragua is down-weighted to contribute only 18\% to the regression results, compared to the full 100\% contribution of most observations. However, this year saw Nicaragua severely affected by a tropical cyclone, and the substantial drop in GDP per capita ($-19\%$) may be linked to the storm. Such an event should not be down-weighted when analyzing the economic impact of severe storms, as it provides valuable information for the study.
\clearpage{}
\section{A Permutation Test for Correlations} \label{sec:testcorr}
Even if all panel data  $Y_{\reg,\tim}$ are independent, we will observe some empirical correlation. We now need to judge whether the correlations we observe are larger than what we would expect in the case of no correlation. 

As the correlation structure in the panel is unclear with respect to both indices, $\reg$ and $\tim$, correlation tests based on an idealized asymptotic distribution and independence in one of the dimensions do not seem to be easy to justify. Instead, we use a permutation test, in which the data is compared to itself. 

The mean correlation of one set of correlations is compared to a null distribution. The null distribution is constructed by taking means of the same number of correlations drawn independently from all correlation values and their negative.

For example, consider correlations of the economic growth rate between years $\tim, \tim\pr\in\Tim$, i.e., the Pearson correlation coefficient $\rho_{\tim, \tim\pr}$ of $(Y_{\reg, \tim})_{\reg\in\Reg_{\tim,\tim\pr}}$ and $(Y_{\reg, \tim\pr})_{\reg\in\Reg_{\tim,\tim\pr}}$, where $\Reg_{\tim,\tim\pr}$ are the countries $\reg$ with $(\reg, \tim), (\reg, \tim\pr) \in \mc I$.
We want to know, if the year-pairs in the set $\mc H_1 \subseteq \Tim \times \Tim$, e.g., consecutive years $(\tim, \tim+1)$, are more correlated on average than other year-pairs. We calculate their mean correlation $\bar\rho_1 := \frac{1}{\#\mc H_1} \sum_{(\tim_1 \tim\pr_1)\in\mc H_1} \rho_{\tim_1, \tim\pr_1}$. Then we repeatedly draw $\#\mc H_1$ values independently from the set $\{\sigma\rho_{\tim_0, \tim\pr_0} | \sigma\in\{-1, 1\}, \tim_0, \tim\pr_0 \in \Tim\}$ and take their mean. The resulting mean values are used to create an empirical distribution function $\hat F$. The $p$-value of this test is then calculated as $p = \hat F(\bar\rho_1)$.

This method is rather conservative as any atypically large correlations (in absolute value) broadens the range that is considered typical. Thus, we will likely not detect all atypical correlations with this test.
\clearpage
\section{On Temporal Stability of Regression Coefficients}\label{sec:tempparab}

\begin{figure}[!h]
    \includegraphics[width=\textwidth]{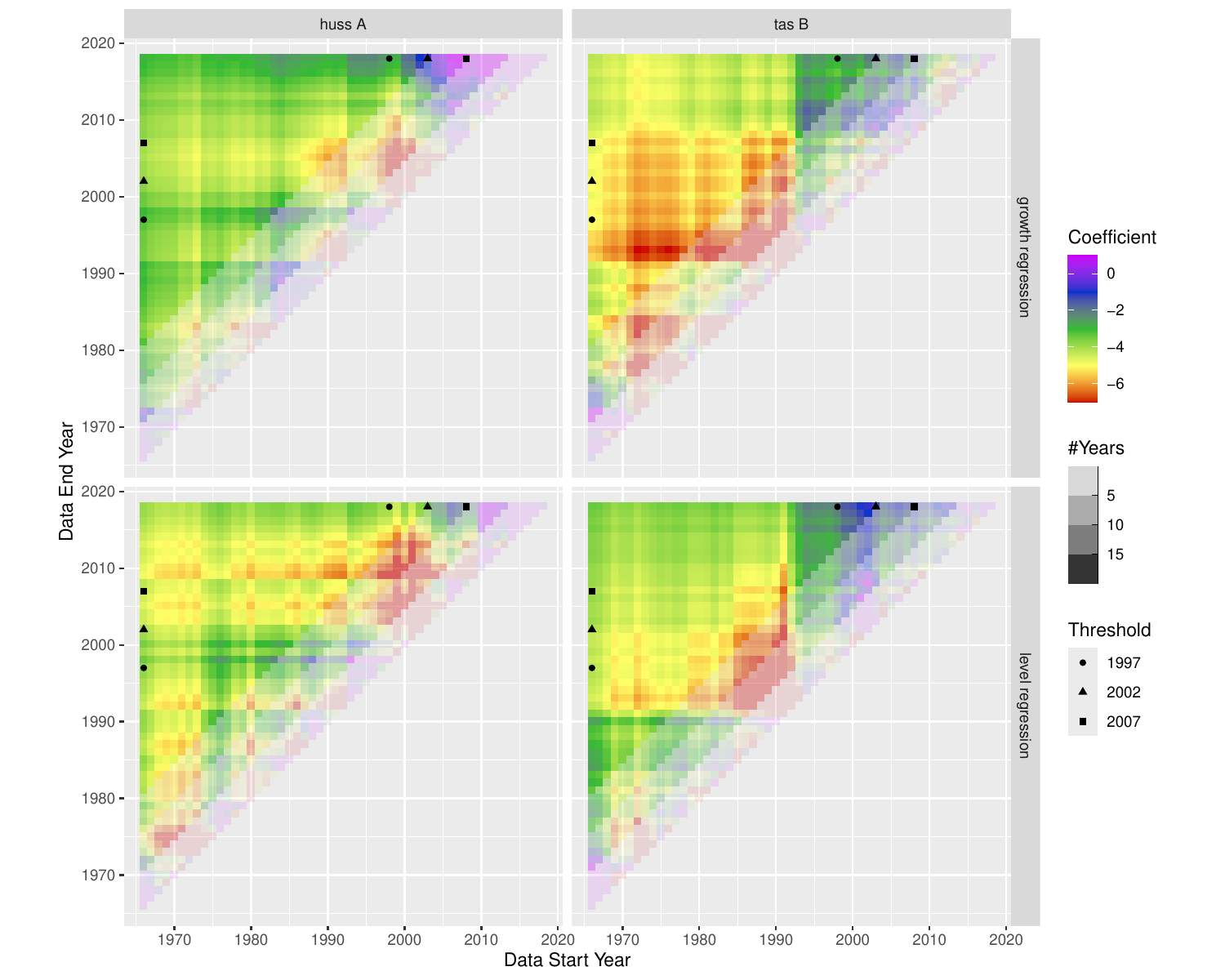}
    \caption{Coefficient values for regression models fitted on cleaned ISIMIP economic data with Kernel controls. Each colored pixel represents the OLS coefficient value for one predictor ($\mathsf{huss}_{\mathsf{max},-1}^{2}$ and $\Delta\mathsf{huss}_{\mathsf{max},-1}^{2}$, respectively, for model huss A, and $\mathsf{tas}_{}^{2}$ and $\Delta\mathsf{tas}_{}^{2}$, respectively, for model tas B, see Table 6 in the main article) calculated on a time period given by the location of the pixel. Note that for short time periods, the coefficient values are not reliable (marked with transparent colors) as too little data is used. The scale of the coefficient values is shown in standard deviations of the respective coefficient estimated with all data and the Driscoll--Kraay method with $8$ lags. In our variable selection study, we split the data into train data and test data using three different thresholds. These are here marked with a black circle, triangle and square, respectively. The symbol in the left most column marks the coefficient obtained from the training data. The corresponding symbol in the top row marks the optimal coefficient for the test data.}
    \label{fig:tempstabi:coef}
\end{figure}
\clearpage
\section{Test Set Significance}\label{sec:tss}
We compare a method's output on the test set with the distribution of an isotropic normal random vector of optimal variance. The variance is chosen depending on a level $p\in[0,1]$ so that the $p$-quantile of the (relative) error is as low as possible. Thus, this null-method chooses a random direction independent of predictors but knows the right order of magnitude to obtain small errors at the level $p$. Next we define this procedure formally.

\begin{figure}
    \includegraphics[width=\textwidth]{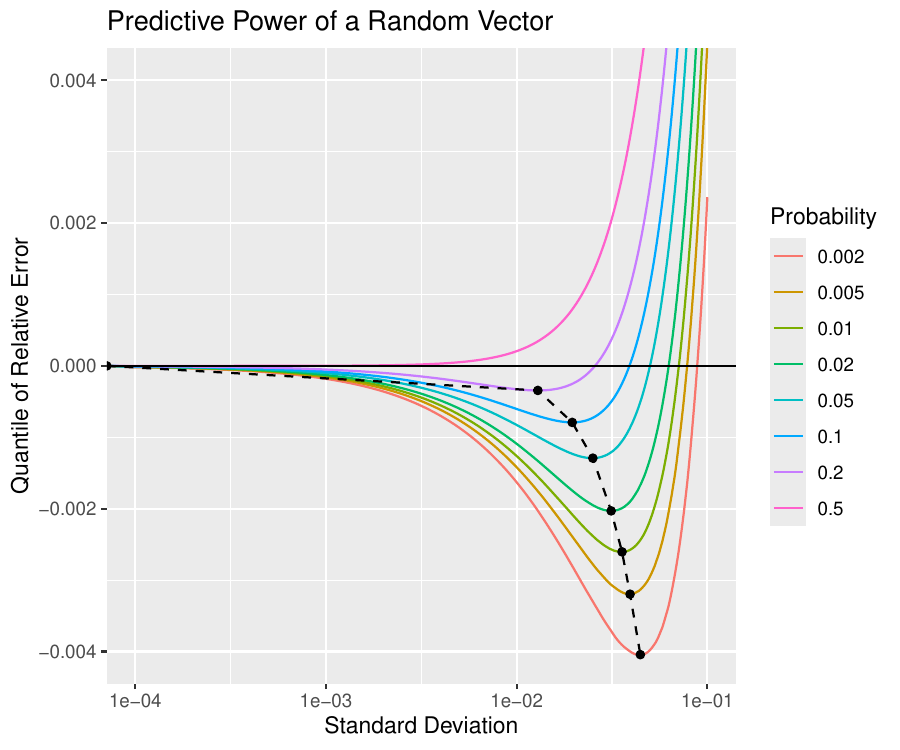}
    \caption{On the controlled dataset, the $p$-quantile of the relative error of an isotropic normal random vector depending on its standard deviation for different values of $p$. In this example, we control the cleaned ISIMIP data with year fixed effects and local linear time trends (Epanechnikov kernel) on show error on the test set after 2007. Each colored line represents a function  $\sigma \mapsto \mathbf{Q}_p[\Vert y - \sigma N\Vert^2]/\Vert y\Vert^2 - 1$ with a different value of $p$. The respective minimum is marked with a black dot. The probability error function $\mathsf{PEF}$ is the mapping from $p$ (indicated by color) to the quantile value at the minimum, i.e., the vertical location of the black dots. As the direction of the random prediction vector is chosen randomly, its median error ($\mathbf{Q}_{0.5}[\Vert y - \sigma N\Vert^2]$) cannot be lower than the trivial $0$-prediction error ($\Vert y - 0\Vert^2 = \Vert y \Vert^2$). But, e.g., in $10\%$ of cases, the error is indeed lower (blue line, $\mathbf{Q}_{0.1}[\Vert y - \sigma N\Vert^2] < \Vert y \Vert^2$) if the scale of the random vector ($\sigma$) is not too large (here $\sigma\lessapprox0.04$ with the lowest value at $\sigma \approx 0.02$).}
    \label{fig:pef}
\end{figure}

Let $y \in \R^m$ be the vector of all response values in the test set. Let $N \sim \mc N(0, I_m)$ be an $m$-dimensional standard normal vector. 
Define the probability error function as the mapping from a probability $p$ to the $p$-quantile of relative squared error of the isotropic normal random vector $\sigma N$ with optimal $\sigma$, i.e.,
\begin{equation}\label{eq:def:errp}
	\mathsf{PEF} \colon [0,1] \to\R, p \mapsto \min_{\sigma}\frac{\mathbf{Q}_p\left[\Vert y - \sigma N\Vert^2\right]-\Vert y\Vert^2}{\Vert y\Vert^2}
	\,,
\end{equation}
where for any real-valued random variable $X$, $ \mathbf{Q}_p[X]$ is a $p$-quantile of $X$, i.e., a value such that
\begin{equation}
	\mathbf{P}(X \leq  \mathbf{Q}_p[X]) \geq p \qquad\text{and}\qquad \mathbf{P}(X \geq  \mathbf{Q}_p[X]) \geq 1-p
	\,.
\end{equation}
The probability error function is monotonically increasing on $[0,\frac12]$ with $\mathsf{PEF}(0) = -1$ and $\mathsf{PEF}(p) = 0$ for $p \geq\frac12$ and, thus, can be inverted. We denote the inverse by $\mathsf{PEF}^{-1}\colon[-1, 0] \to [0, \frac12]$. Furthermore, we set $\mathsf{PEF}^{-1}(a) = 1$ for $a > 0$.
For a given method's predictions $\hat y$ on the test set, we output the pseudo-$p$-value
\begin{equation}
	\mathsf{PEF}^{-1}(\Vert y - \hat y\Vert^2)
	\,.
\end{equation}
In practice, we approximate the minimum in \eqref{eq:def:errp} by a grid search of $10^3$ values on a logarithmic scale for $\sigma$ and the quantile function by $10^5$ samples of $N$. A simulation result for the $\ms{PEF}$ is shown in \cref{fig:pef}.

\clearpage
\section{Comparison to \citet{newell_gdp-temperature_2021}}\label{sec:newell}
To the best of our knowledge, our study is only the second work that uses out-of-sample testing to evaluate global climate econometric panel regression models. The first study is \citet{newell_gdp-temperature_2021}. 
Other than out-of-sample testing for climate econometric panel data regression, there are not many aspects this studies has in common with our work. Still, we want to describe the similarities and differences specifically in the out-of-sample testing methodology and results, which we deem relevant for future work.

\citet{newell_gdp-temperature_2021} remove fixed effects before data is split into train as test set, but estimate the country-specific time trends on the training data and use it to extrapolate these polynomial trends to the test data. This may be problematic as polynomial extrapolation---in particular for higher degrees---can only be predictive under very strong assumptions. Furthermore, as a polynomial is fitted for each country there may be an overfitting issue. In their results, models with extrapolated time trends have much larger errors on the test set (their figure 1), with higher errors for here polynomial degrees in the time trends. Because of these results, they dismiss all models with time trend controls.

We instead remove all control variables---fixed effects and country-wise time-trends---from the data before the train-test split. Furthermore, we cannot and do not directly rate models with different control designs based on their test errors. Rather, we evaluate temporal and spatial correlation in the controlled data to allow a comparison of different control designs. This comparison is thus separate from and independent of any downstream model comparisons that evaluate different climate predictor variables.

We are interested the influence of climate predictors on economic output rather than the time trends, which we use removing temporal confounding effects to increase the plausibility of a causal relationship in our results. Furthermore, the Frisch–Waugh–Lovell theorem justifies the two-step procedure of first removing controls and then fitting the predictor variables, which results in the same estimated coefficients as when running the full regression in one step. 

\clearpage
\section{Comparison of Model Selection Procedures via Test Error}\label{sec:selection:compare}

For all six model selection procedures, we collect the mean squared error (MSE) of the selected models on the test set (out-of-sample) in the six tables \cref{tbl:testmethods:growth:1997,tbl:testmethods:growth:2002,tbl:testmethods:growth:2007,tbl:testmethods:level:1997,tbl:testmethods:level:2002,tbl:testmethods:level:2007}. Each table shows the results for one combination of a regression design (growth or level) and a threshold year for the split into training and test data (1997, 2002, 2007). 

Results are shown in each table for four different economic datasets (ISIMIP and World Bank, each cleaned or raw). Furthermore, we show results for eight different control designs: None (no controls), Year (year fixed effects $\alpha_\tim$), YearPoly0 (year and county fixed effects $\alpha_\tim + \alpha_{\reg, 0}$), YearPoly1 (year and county fixed effects with country-wise linear time trends $\alpha_\tim + \alpha_{\reg,0} + \alpha_{\reg,1} t$), YearPoly2 (year and county fixed effects with country-wise quadratic time trends $\alpha_\tim + \alpha_{\reg,0} + \alpha_{\reg,1} t + \alpha_{\reg,2} t^2$), YearPoly3 (year and county fixed effects with country-wise cubic time trends $\alpha_\tim + \alpha_{\reg,0} + \alpha_{\reg,1} t + \alpha_{\reg,2} t^2 + \alpha_{\reg,3} t^3$), Window (application of controls $\alpha_\tim + \alpha_{\reg,0} + \alpha_{\reg,1} t$ in a rolling window over time with uniform weights), Kernel (application of controls $\alpha_\tim + \alpha_{\reg,0} + \alpha_{\reg,1} t$ in a rolling window over time with quadratically decreasing weights). 

The MSE is shown in percent and relative to the MSE of the null-model, which has no predictors (it predicts constant $0$ growth in the controlled data). Thus, the lowest possible value is $-100$ meaning perfect prediction of the test set. A value above $0$ means that the selected model is worse than the trivial baseline. Additionally we show a $p$-value in percent, which is described in \cref{sec:tss}. If the MSE is negative (i.e., the selected model has predictive power on the test set) but the $p$-value is high (say, more than 10\%) then it remains questionable whether or not the selected model has predictive power as a similar test error could be obtained by random guessing. 

\begin{table}[h!]
	\include{tbl/09_together_growth_1997.tex}
	\caption{See \cref{sec:selection:compare} for a description.}
	\label{tbl:testmethods:growth:1997}
\end{table}

\begin{table}[h!]
	\include{tbl/09_together_growth_2002.tex}
	\caption{See \cref{sec:selection:compare} for a description.}
	\label{tbl:testmethods:growth:2002}
\end{table}

\begin{table}[h!]
	\include{tbl/09_together_growth_2007.tex}
	\caption{See \cref{sec:selection:compare} for a description.}
	\label{tbl:testmethods:growth:2007}
\end{table}

\begin{table}[h!]
	\include{tbl/09_together_level_1997.tex}
	\caption{See \cref{sec:selection:compare} for a description.}
	\label{tbl:testmethods:level:1997}
\end{table}

\begin{table}[h!]
	\include{tbl/09_together_level_2002.tex}
	\caption{See \cref{sec:selection:compare} for a description.}
	\label{tbl:testmethods:level:2002}
\end{table}

\begin{table}[h!]
	\include{tbl/09_together_level_2007.tex}
	\caption{See \cref{sec:selection:compare} for a description.}
	\label{tbl:testmethods:level:2007}
\end{table}

\clearpage
\section{Comparison of Variable Selection for Full Training Data}

\begin{table}[h!]
	\include{tbl/09_together_variableSelection_growth_2018.tex}
	\caption{Variable re-selection rate in \% from bootstrap on growth regression models using the all data, i.e., until 2018.}
	\label{tbl:test:selection:main:growth}
\end{table}

\begin{table}[h!]
	\include{tbl/09_together_variableSelection_level_2018.tex}
	\caption{Variable re-selection rate in \% from bootstrap on level regression models using the all data, i.e., until 2018.}
	\label{tbl:test:selection:main:level}
\end{table}

\clearpage
\section{Variable Selection Results}\label{sec:regresults}

For each model selection method that explicitly selects variables (Forward Selection, LASSO, Elastic Net, Pair Selection), we show six tables with different regression designs (level vs growth) and last year of training (1997, 2003, 2007). Each table contains results for all combinations of economic datasets and controls as in \cref{sec:selection:compare}. The columns MSE\% and p\% show the result of the selected model on the test set as in \cref{sec:selection:compare}. We first use the given training data for variable selection and note the number of selected variables in the column $\#$. The most commonly selected predictors are shown in the following columns with the number of not explicitly shown variables in the last column $+\#$. The numbers in the predictor columns are the re-selection rate in percent from a block bootstrap keeping all data of the same year in the same block. These numbers indicate how robust the selection of the given predictor is with high numbers (say $\geq 90$) marking robust choices.

\clearpage
\subsection{Forward Regression Results}

\begin{table}[h!]
	\include{tbl/09_forward_growth_1997.tex}
	\caption{Detailed description in \cref{sec:regresults}.}
	\label{tbl:regres:forward:growth:1997}
\end{table}

\begin{table}[h!]
	\include{tbl/09_forward_growth_2002.tex}
	\caption{Detailed description in \cref{sec:regresults}.}
	\label{tbl:regres:forward:growth:2002}
\end{table}

\begin{table}[h!]
	\include{tbl/09_forward_growth_2007.tex}
	\caption{Detailed description in \cref{sec:regresults}.}
	\label{tbl:regres:forward:growth:2007}
\end{table}

\begin{table}[h!]
	\include{tbl/09_forward_growth_2018.tex}
	\caption{Detailed description in \cref{sec:regresults}.}
	\label{tbl:regres:forward:growth:2018}
\end{table}

\begin{table}[h!]
	\include{tbl/09_forward_level_1997.tex}
	\caption{Detailed description in \cref{sec:regresults}.}
	\label{tbl:regres:forward:level:1997}
\end{table}

\begin{table}[h!]
	\include{tbl/09_forward_level_2002.tex}
	\caption{Detailed description in \cref{sec:regresults}.}
	\label{tbl:regres:forward:level:2002}
\end{table}

\begin{table}[h!]
	\include{tbl/09_forward_level_2007.tex}
	\caption{Detailed description in \cref{sec:regresults}.}
	\label{tbl:regres:forward:level:2007}
\end{table}

\begin{table}[h!]
	\include{tbl/09_forward_level_2018.tex}
	\caption{Detailed description in \cref{sec:regresults}.}
	\label{tbl:regres:forward:level:2018}
\end{table}

\clearpage
\subsection{Lasso Results}

\begin{table}[h!]
	\include{tbl/09_lasso_growth_1997.tex}
	\caption{Detailed description in \cref{sec:regresults}.}
	\label{tbl:regres:lasso:growth:1997}
\end{table}

\begin{table}[h!]
	\include{tbl/09_lasso_growth_2002.tex}
	\caption{Detailed description in \cref{sec:regresults}.}
	\label{tbl:regres:lasso:growth:2002}
\end{table}

\begin{table}[h!]
	\include{tbl/09_lasso_growth_2007.tex}
	\caption{Detailed description in \cref{sec:regresults}.}
	\label{tbl:regres:lasso:growth:2007}
\end{table}

\begin{table}[h!]
	\include{tbl/09_lasso_growth_2018.tex}
	\caption{Detailed description in \cref{sec:regresults}.}
	\label{tbl:regres:lasso:growth:2018}
\end{table}

\begin{table}[h!]
	\include{tbl/09_lasso_level_1997.tex}
	\caption{Detailed description in \cref{sec:regresults}.}
	\label{tbl:regres:lasso:level:1997}
\end{table}

\begin{table}[h!]
	\include{tbl/09_lasso_level_2002.tex}
	\caption{Detailed description in \cref{sec:regresults}.}
	\label{tbl:regres:lasso:level:2002}
\end{table}

\begin{table}[h!]
	\include{tbl/09_lasso_level_2007.tex}
	\caption{Detailed description in \cref{sec:regresults}.}
	\label{tbl:regres:lasso:level:2007}
\end{table}

\begin{table}[h!]
	\include{tbl/09_lasso_level_2018.tex}
	\caption{Detailed description in \cref{sec:regresults}.}
	\label{tbl:regres:lasso:level:2018}
\end{table}

\clearpage
\subsection{Elastic Net Results}

\begin{table}[h!]
	\include{tbl/09_elastic_growth_1997.tex}
	\caption{Detailed description in \cref{sec:regresults}.}
	\label{tbl:regres:elastic:growth:1997}
\end{table}

\begin{table}[h!]
	\include{tbl/09_elastic_growth_2002.tex}
	\caption{Detailed description in \cref{sec:regresults}.}
	\label{tbl:regres:elastic:growth:2002}
\end{table}

\begin{table}[h!]
	\include{tbl/09_elastic_growth_2007.tex}
	\caption{Detailed description in \cref{sec:regresults}.}
	\label{tbl:regres:elastic:growth:2007}
\end{table}

\begin{table}[h!]
	\include{tbl/09_elastic_growth_2018.tex}
	\caption{Detailed description in \cref{sec:regresults}.}
	\label{tbl:regres:elastic:growth:2018}
\end{table}

\begin{table}[h!]
	\include{tbl/09_elastic_level_1997.tex}
	\caption{Detailed description in \cref{sec:regresults}.}
	\label{tbl:regres:elastic:level:1997}
\end{table}

\begin{table}[h!]
	\include{tbl/09_elastic_level_2002.tex}
	\caption{Detailed description in \cref{sec:regresults}.}
	\label{tbl:regres:elastic:level:2002}
\end{table}

\begin{table}[h!]
	\include{tbl/09_elastic_level_2007.tex}
	\caption{Detailed description in \cref{sec:regresults}.}
	\label{tbl:regres:elastic:level:2007}
\end{table}

\begin{table}[h!]
	\include{tbl/09_elastic_level_2018.tex}
	\caption{Detailed description in \cref{sec:regresults}.}
	\label{tbl:regres:elastic:level:2018}
\end{table}

\clearpage
\subsection{Pair Selection Results}

\begin{table}[h!]
	\include{tbl/09_subset2_growth_1997.tex}
	\caption{Detailed description in \cref{sec:regresults}.}
	\label{tbl:regres:subset2:growth:1997}
\end{table}

\begin{table}[h!]
	\include{tbl/09_subset2_growth_2002.tex}
	\caption{Detailed description in \cref{sec:regresults}.}
	\label{tbl:regres:subset2:growth:2002}
\end{table}

\begin{table}[h!]
	\include{tbl/09_subset2_growth_2007.tex}
	\caption{Detailed description in \cref{sec:regresults}.}
	\label{tbl:regres:subset2:growth:2007}
\end{table}

\begin{table}[h!]
	\include{tbl/09_subset2_growth_2018.tex}
	\caption{Detailed description in \cref{sec:regresults}.}
	\label{tbl:regres:subset2:growth:2018}
\end{table}

\begin{table}[h!]
	\include{tbl/09_subset2_level_1997.tex}
	\caption{Detailed description in \cref{sec:regresults}.}
	\label{tbl:regres:subset2:level:1997}
\end{table}

\begin{table}[h!]
	\include{tbl/09_subset2_level_2002.tex}
	\caption{Detailed description in \cref{sec:regresults}.}
	\label{tbl:regres:subset2:level:2002}
\end{table}

\begin{table}[h!]
	\include{tbl/09_subset2_level_2007.tex}
	\caption{Detailed description in \cref{sec:regresults}.}
	\label{tbl:regres:subset2:level:2007}
\end{table}

\begin{table}[h!]
	\include{tbl/09_subset2_level_2018.tex}
	\caption{Detailed description in \cref{sec:regresults}.}
	\label{tbl:regres:subset2:level:2018}
\end{table}

%% file: tbl/01_DOSE_Spain_Meta.tex
\begin{center}
\fontsize{8.0pt}{10pt}\selectfont
\fontfamily{phv}\selectfont
\renewcommand{\arraystretch}{1.05}
\setlength{\tabcolsep}{0.3em}
\rowcolors{2}{gray!20}{white}
\begin{tabular}{llll}
\toprule
GID\_0 & GID\_1 & Country & Region \\ 
\midrule\addlinespace[2.5pt]
ESP & ESP.9\_1 & Spain & Comunidad Foral de Navarra \\ 
ESP & ESP.10\_1 & Spain & Comunidad Valenciana \\ 
ESP & ESP.8\_1 & Spain & Comunidad de Madrid \\ 
\bottomrule
\end{tabular}
\end{center}